\documentclass[a4paper,11pt]{article}

\usepackage{jheppub} 

\usepackage{amsmath}
\usepackage[utf8]{inputenc}
\usepackage{amssymb}
\usepackage{amsfonts}
\usepackage{amsbsy}
\usepackage{url}
\usepackage{enumerate}
\usepackage{caption}
\usepackage{subcaption}
\usepackage{color}
\usepackage{bm}
\usepackage{bbm}
\usepackage{comment}
\usepackage{multirow,bigdelim}
\usepackage[head1]{DotArrow}
\usepackage[normalem]{ulem}

\usepackage{mathrsfs}

\newcommand{\be}{\begin{eqnarray}}
\newcommand{\ee}{\end{eqnarray}}

\makeatletter
\@addtoreset{equation}{section}
\makeatother

\newcommand{\NF}{N_{\rm F}}
\newcommand{\NC}{N_{\rm C}}
\newcommand{\beq}{\begin{eqnarray}}
\newcommand{\eeq}{\end{eqnarray}}

\newcommand{\pd}{{\partial}}

\def\({\left(}
\def\){\right)}
\def\[{\left[}
\def\]{\right]}

\def\<{\left\langle}
\def\>{\right\rangle}

\newcommand{\n}{{\bf n}}
\newcommand{\A}{\bm{A}}

\graphicspath{{./figs/}}
\newbox{\ORCIDicon}

\title{
Chiral Magnets from String Theory
}

\author[a,b]{Yuki Amari,\,\href{https://orcid.org/0000-0002-3900-8507}{\usebox{\ORCIDicon}}}
\emailAdd{amari.yuki@keio.jp}
\affiliation[a]{Research and Education Center for Natural Sciences, Keio University, 4-1-1 Hiyoshi, Yokohama, Kanagawa 223-8521, Japan}
\author[a,b,c]{and Muneto Nitta,\,\href{https://orcid.org/0000-0002-3851-9305	}{\usebox{\ORCIDicon}}}
\emailAdd{nitta@phys-h.keio.ac.jp}
\affiliation[d]{Department of Physics, Keio University, 4-1-1 Hiyoshi, Yokohama, Kanagawa 223-8521, Japan}
\affiliation[e]{
International Institute for Sustainability with Knotted Chiral Meta Matter(SKCM$^2$), Hiroshima University, 1-3-2 Kagamiyama, Higashi-Hiroshima, Hiroshima 739-8511, Japan
}

\abstract{
Chiral magnets with the Dzyaloshinskii-Moriya (DM) interaction
have received quite an intensive focus in condensed matter physics because of the presence of 
a chiral soliton lattice (CSL), 
an array of magnetic domain walls and anti-domain walls,  and magnetic skyrmions, 
both of which are 
important ingredients in the current nanotechnology. 
In this paper, we realize chiral magnets in type-IIA/B string theory by 
using the Hanany-Witten brane configuration 
(consisting of D3, D5 and NS5-branes)
and the fractional D2 and D6 branes on the Eguchi-Hanson manifold. 
In the both cases,   
we put constant non-Abelian magnetic fluxes on 
higher dimensional (flavor) 
D-branes, turning them into 
magnetized D-branes.
The $O(3)$ sigma model with 
an easy-axis or easy-plane potential and 
the DM interaction  
is realized on 
the worldvolume of 
the lower dimensional (color) D-branes. 
The ground state is 
the ferromagnetic (uniform) phase and the color D-brane is straight 
when the DM interaction is small compared with the scalar mass. 
However, when the DM interaction is larger, 
the uniform state is no longer 
stable and the ground state 
is inhomogeneous: the CSL phases and helimagnetic phase.
In this case, 
the color D-brane is no longer straight  but is 
snaky (zigzag) 
when the DM interaction 
is smaller (larger) 
than a critical value. 
A magnetic domain wall 
in the ferromagnetic phase is 
realized as a kinky D-brane. 
We further construct magnetic skyrmions in the ferromagnetic phase, 
realized as D1-branes (fractional D0-branes)
in 
the former (latter) configuration.
We see that the host D2-brane is bent around the position of 
a D0-brane as a magnetic skyrmion.
Finally, we construct, 
in the ferromagnetic phase, 
domain-wall skyrmions, 
that is, composite states of a domain wall and skyrmions, 
and find that the domain wall 
is no longer flat in the vicinity of the skyrmion.
Consequently, 
a kinky D2-brane worldvolume 
is pulled or pushed 
in the vicinity of the D0-brane 
depending on the sign of the skyrmion topological charge.
}

\begin{document}

\maketitle


\section{Introduction}
Recently chiral magnets accompanied with the Dzyaloshinskii-Moriya (DM) interaction
\cite{Dzyaloshinskii,Moriya:1960zz} 
have been paid great attention 
in condensed matter physics, 
because of the presence 
of a chiral soliton lattice (CSL), 
an array of solitons or a pair of 
magnetic domain walls 
and anti-domain walls 
\cite{togawa2012chiral,togawa2016symmetry,KISHINE20151,
PhysRevB.97.184303,PhysRevB.65.064433,Ross:2020orc},  
and magnetic skyrmions 
\cite{Bogdanov:1989,Bogdanov:1995}. 
In a certain parameter region of chiral magnets, 
a CSL is the ground state 
\cite{togawa2012chiral,togawa2016symmetry,KISHINE20151,
PhysRevB.97.184303,PhysRevB.65.064433,Ross:2020orc}
where the energy of a single soliton is negative, and one dimensional modulated states have lower energy than a uniform state (ferromagnetic phase). 
On the other hand,
magnetic skyrmions \cite{Bogdanov:1989,Bogdanov:1995} have received quite intensive focus due to their realizations  
in a form of skyrmion lattices in the laboratory in chiral magnets~\cite{doi:10.1126/science.1166767,doi:10.1038/nature09124,doi:10.1038/nphys2045}
(see also Refs.~\cite{Rossler:2006,Han:2010by,Lin:2014ada,Ross:2022vsa})
in addition to noncentrosymmetric magnets~\cite{Kurumaji2019,Hirschberger2019,Khanh2020,Yasui2020},  
and their possible applications as components for data storage with low energy consumption
\cite{doi:10.1038/nnano.2013.29}, 
see Ref.~\cite{Nagaosa2013} for a review. 
As a combination of a domain wall and skyrmion, 
domain-wall skyrmions were first proposed 
in quantum field theory \cite{Nitta:2012xq,Kobayashi:2013ju} (see also Refs.~\cite{Auzzi:2006ju,Jennings:2013aea,Bychkov:2016cwc})\footnote{
The term 
``domain wall skyrmions'' was first introduced in Ref.~\cite{Eto:2005cc} 
in which Yang-Mills instantons 
in the bulk are 
3D skyrmions inside a domain wall. 
Domain-wall skyrmions in 3+1 dimensions for which 3D skyrmions 
are 2D skyrmions on a domain wall 
were proposed in field theory 
\cite{Nitta:2012wi,Nitta:2012rq,
Gudnason:2014nba,Gudnason:2014hsa,Eto:2015uqa} and are 
recently realized in QCD 
\cite{Eto:2023lyo,Eto:2023wul,Eto:2023tuu}.
} 
and have been recently 
 observed  experimentally in chiral magnets 
\cite{PhysRevB.102.094402,Nagase:2020imn,Yang:2021} (see also \cite{Kim:2017lsi}).
A first step at treating chiral magnetic domain walls theoretically is given in Refs.~\cite{PhysRevB.99.184412,KBRBSK,Ross:2022vsa}. 
Apart from domain walls and skyrmions, 
a lot of studies have been devoted to 
various topological objects 
such as 
monopoles \cite{tanigaki2015,fujishiro2019topological}, 
Hopfions \cite{Sutcliffe:2018vcb} 
and 
instantons \cite{Hongo:2019nfr}, 
see Ref.~\cite{GOBEL20211} as a review.

In spite of such great interests in condensed matter physics and materials science,
chiral magnets were not paid 
much attention in high energy physics, 
since the DM interaction is not Lorentz invariant.  
One exception would be the finding of 
Bogomolnyi-Prasad-Sommerfield (BPS) magnetic skyrmions \cite{Barton-Singer:2018dlh,Schroers:2019hhe,Ross:2020hsw}.

In this paper, 
we give a possible realization of chiral magnets in string theory. 
One of the key points is the DM interaction realized as a background $SU(2)$ gauge field
\cite{Barton-Singer:2018dlh,Schroers:2019hhe}.
The other is the Hanany-Witten brane configuration 
\cite{Hanany:1996ie,Giveon:1998sr} 
and magnetized D-branes \cite{Bachas:1995ik,Berkooz:1996km,Blumenhagen:2000wh,Angelantonj:2000hi,
Cremades:2004wa,Kikuchi:2023awm,Abe:2021uxb}.
We first formulate 
the $O(3)$ model, or
the ${\mathbb C}P^1$ model, 
with the DM interaction 
in terms of 
a $U(1) \times SU(2)$ 
gauge theory. 
The $U(1)$ gauge symmetry is 
an auxiliary field 
and the $SU(2)$ gauge symmetry 
is a background gauge field.
Next, we embed this gauge theory 
into two kinds of D-brane configurations: 
one is the Hanany-Witten type brane configuration 
in type-IIB string theory, 
composed of two NS5-branes stretched by D3-branes 
and orthogonal D5-branes 
\cite{Hanany:1996ie,Giveon:1998sr}, 
and the other is a D2-D6-brane bound state 
on the Eguchi-Hanson manifold in type-IIA string theory. 
A chiral magnet is realized on 
the worldvolume of 
the lower dimensional D-branes, 
D3-branes in the former 
and D2-branes in the latter. 

The potential is classified into 
an easy-axis or easy-plane potential.
The ground state is either a ferromagnetic\footnote{
The $O(3)$ model with the Lorentz invariant kinetic term that we are considering 
is relevant to rather antiferromagnets than ferromagnets, 
and thus it should be called antiferromagnetic strictly speaking. 
However, in this paper, we call the uniform state ``ferromagnetic'' for simplicity.
} (uniform) phase 
or inhomogeneous phases when the DM interaction is smaller 
or larger than a critical value, respectively.
The inhomogeneous ground states are further classified 
into three cases: 
two kinds of the CSL phases with the easy-axis and easy-plane potentials 
and a linearly modulated phase (with no potential) 
at the boundary of 
these two CSL phases. 
In these different phases, the lower dimensional color 
D-branes behave differently in the brane configurations.
In the ferromagnetic phase, 
the color D-brane is straight 
in the vacuum, 
and 
a magnetic domain wall as an excited state 
 is 
realized as a kinky D-brane 
\cite{Lambert:1999ix,Eto:2004vy,Eto:2006mz,Eto:2007aw,Misumi:2014bsa}.
In the CSL phase with the easy-axis potential, 
the color D-brane is {\it snaky} 
(an array of kinky D-branes and anti-kinky D-branes) between the two separated 
flavor D-branes.
On the other hand, 
in the CSL phase with the easy-plane potential, 
the color D-brane is rather {\it zigzag} 
between the two flavor D-branes. 
In this case, 
each separated (anti-)domain wall is nontopological.
Between these two CFL phases, 
the ground state is helimagnetic 
in which the color D-brane 
modulates as a sine function.

We further study magnetic skyrmions. 
They are realized as D1-branes 
in the Hanany-Witten type brane configuration 
as an analog of vortices 
\cite{Hanany:2003hp}.
On the other hand, 
in the D2-D6-ALE system,
they are 
fractional D0-branes, that is, D2-branes 
two directions of whose worldvolumes wrap around 
the $S^2$ cycle blowing up the ${\mathbb Z}_2$ orbifold singularity.
We show that the host D2(D3)-brane is bent at 
the position of 
a D0(D1)-brane as a magnetic skyrmion.
Finally, we construct domain-wall skyrmions 
\cite{Nitta:2012xq,Kobayashi:2013ju,Ross:2022vsa,Nitta:2022ahj} in the ferromagnetic phase. 
Magnetic (anti-)skyrmions are realized as (anti-) sine-Gordon solitons in a magnetic domain wall, 
whose worldvolume theory 
is the sine-Gordon model 
with a potential term coming from 
the DM interaction \cite{Ross:2022vsa}.
The domain-wall worldvolume 
is no longer flat in the vicinity of the skyrmion \cite{KBRBSK}.
Consequently, the color D-brane worldvolume, 
forming a kink for the magnetic domain wall, 
is pulled 
in the vicinity of the (anti-)skyrmion to either side of the domain 
wall depending on whether it is a skyrmion or an anti-skyrmion.

This paper is organized as follows.
In Sec.~\ref{sec:gauge-theory} 
we formulate the $O(3)$ model, or the ${\mathbb C}P^1$ model, with the DM interaction as a $U(1) \times SU(2)$ gauge theory.
In Sec.~\ref{sec:dbrane}, we present 
D-brane configurations for the chiral magnets, 
the Hanany-Witten brane configuration 
and the D2-D6-ALE system.
We put a non-Abelian 
magnetic flux breaking the $SU(2)$ flavor symmetry 
into $U(1)$, making 
the D5-branes in the former 
or the D6-branes in the latter magnetized.
In Sec.~\ref{sec:wall-CSL},  
we construct a magnetic domain wall 
and the CSL ground states 
forming a snaky D-brane 
and a zigzag D-brane in the former 
and the latter brane configurations,
respectively. 
In Sec.~\ref{sec:skyrmions}, 
we construct magnetic skyrmions 
and domain-wall skyrmions, 
and discuss their D-brane configurations.
Sec.~\ref{sec:summary} is devoted 
to a summary and discussion.
In Appendix.~\ref{sec:DM-from-BGG}, 
we give a derivation of the DM interaction 
from the gauged linear sigma model.

\section{Chiral Magnets as Gauge Theory}
\label{sec:gauge-theory}

In this section, 
we formulate the $O(3)$ model describing Heisenberg magnets as a gauge theory for the cases without the DM term in Subsec.~\ref{sec:Heisenberg-gauge} and with the DM term 
in Subsec.~\ref{sec:chiral-gauge}.

\subsection{Heisenberg magnets from gauge theory} \label{sec:Heisenberg-gauge}
We start with a $U(1)$ gauge theory 
with a gauge field 
$a_{\mu}$ coupled with 
complex scalar fields 
written as $\Phi^T = (\Phi_1,\Phi_2)$ 
and a real scalar field $\Sigma$.
The Lagrangian is 
\begin{eqnarray}
&& {\cal L} = -\frac{1}{4 g^2} F_{\mu\nu}F^{\mu\nu} 
 + \frac{1}{g^2} (\partial_{\mu} \Sigma)^2 
 + 2|D_{\mu} \Phi|^2 - V\\
&& V = {g^2\over 2} (\Phi^\dagger\Phi -v^2)^2 
 + \Phi^\dagger(\Sigma {\bf 1}_2 - M)^2 \Phi 
\end{eqnarray}
with the gauge coupling  $g$, 
the vacuum expectation value (VEV) $v$ of $\Phi$, and 
the covariant derivative $D_{\mu} \Phi 
   = (\partial_\mu - i a_{\mu}) \Phi$.
Here, $M$ is a mass matrix of $\Phi$ given by 
$M={\rm diag}(m,-m)$ with a constant $m$, 
where the overall diagonal constant can be eliminated 
by a redefinition of $\Sigma$.
This can be made ${\cal N}=2$ supersymmetric(SUSY) 
with eight supercharges by adding 
a complex scalar field 
$\tilde \Phi$, 
and fermionic superpartners \cite{Eto:2006pg}:
$(\Phi,\tilde \Phi)$
with fermionic superpartners called Higgsinos are 
hypermultiplets, 
and $(a_{\mu}, \Sigma)$ 
with a fermionic superpartner called a gaugino 
is a gauge or vector multiplet.

In the strong  coupling limit $g^2 \to \infty$, 
the kinetic terms of 
$a_{\mu}$ and $\Sigma$ 
disappear and they become auxiliary fields, which can be eliminated 
by their equations of motion:
\begin{eqnarray}
a_{\mu} = 
\frac{i}{2v^2} 
(\partial_{\mu}\Phi^\dagger \cdot \Phi - \Phi^\dagger \partial_{\mu}\Phi), \quad
    \Sigma  
    =\frac{1}{v^2}\Phi^\dagger M \Phi = \frac{mn_3}{v^2} 
    \ .
    \label{eq:elimination_auxiliary}
\end{eqnarray}
Then, the model reduces to the ${\mathbb C}P^1$ model with a potential term.
By rewriting 
\begin{eqnarray}
    \Phi^T = v (1,u)/\sqrt{1 + |u|^2}
\end{eqnarray} 
with a 
complex projective coordinate $u$, 
the Lagrangian becomes 
\begin{eqnarray}
&& {\cal L} =
 {2 \partial_{\mu} u^* \partial^{\mu} u - 4 m^2 |u|^2 
  \over (1 + |u|^2)^2} \ .
  \label{eq:CP1}
\end{eqnarray}
We have set $v=1$ for simplicity.
This model is known as the massive 
${\mathbb C}P^1$ model 
with the potential term
\begin{eqnarray}
V= {4 m^2 |u|^2 
  \over (1 + |u|^2)^2} 
\end{eqnarray}
which is 
the Killing vector squared 
corresponding to the isometry generated by $\sigma_z$, 
and 
admits two discrete vacua $u=0$ and $u=\infty$. 
This construction is known as 
a K\"{a}hler quotient.

Introducing a three-vector of scalar fields by
\begin{eqnarray}
\n=\Phi^\dagger \bm{\sigma}\Phi
\end{eqnarray}
with the Pauli matrices $\bm{\sigma}$, 
the Lagrangian can be rewritten in the form of the $O(3)$ model:
\begin{eqnarray}
 {\cal L} = \frac{1}{2} \partial_{\mu}{\bf n} \cdot \partial^{\mu}{\bf n} 
 - m^2(1-n_3^2) ,
 \quad {\bf n}^2 =1.
\end{eqnarray}
This model is known as a continuum limit of the (anti-ferromagnetic)  Heisenberg model with 
  an easy-axis potential 
 $V = m^2(1-n_3^2)$.

\subsection{Dzyaloshinskii-Moriya interaction}
\label{sec:chiral-gauge}
Now we are ready to introduce 
the DM term. 
We can achieve this by gauging the $SU(2)$ flavor symmetry with a background gauge field
\cite{Barton-Singer:2018dlh,Schroers:2019hhe}.
The Lagrangian is now a $U(1) \times SU(2)$ gauge theory
\begin{eqnarray}
&& {\cal L} = -\frac{1}{4 g^2} F_{\mu\nu}F^{\mu\nu} 
 + \frac{1}{g^2} (\partial_{\mu} \Sigma)^2 
 + 2 |{\cal D}_{\mu} \Phi|^2 - V
 \label{eq:U2_gauge_theory}
\end{eqnarray}
with a $U(1) \times SU(2)$ covariant derivative  
\begin{eqnarray}
   {\cal D}_{\mu} \Phi 
   = \left(\partial_\mu - i a_{\mu} - \frac{i}{2} A_{\mu}
   \right) \Phi
\end{eqnarray}
and an $SU(2)$ background gauge field 
$A_{\mu} = A_{\mu}^a \sigma_a$.

In the strong gauge coupling limit $g^2 \to \infty$, 
$a_{\mu}$ and $\Sigma$ become auxiliary fields as before.
After eliminating these auxiliary fields $a_{\mu}$ and $\Sigma$ 
as in the previous subsection,
we reach at 
(see Appendix 
\ref{sec:DM-from-BGG}
for a derivation):
\begin{eqnarray}
    {\cal L} =    \frac{1}{2}\left(D_{\mu} \n \right)^{2} 
    -m^{2} (1-n_{3}^{2})
    \label{eq:gauged_sigma_model}
\end{eqnarray}
with $D_{\mu} \n =\partial_{\mu} \n+\A_{\mu} \times \n$.
Here we took the temporal gauge 
$\A_0=(0,0,0)$,
and $\A_{\mu=k} =(A_k^1, A_k^2, A_k^3)$ are spatial components 
of a three vector with the $SU(2)$ adjoint index for which 
the product $\times$ is defined.
The corresponding Hamiltonian density is 
\begin{eqnarray}
 {\cal H} &=& \frac{1}{2}\left(D_{k} \n \right)^{2} 
    + m^{2} (1-n_{3}^{2}) \nonumber\\
  &=& \frac{1}{2} \partial_{k}\n \cdot\partial_{k} \n 
  +\A_{k} \cdot\left( \n \times \partial_{k} \n \right)
  +\frac{1}{2}\left(\A_{k} \times\n\right)^{2} + m^{2} (1- n_{3}^{2}) 
  \label{eq:generalized-DM}
\end{eqnarray}
where the second term gives 
the DM term 
\begin{eqnarray}
  {\cal H}_{\rm DM} \equiv \A_{k} \cdot\left( \n \times \partial_{k} \n \right)   
\end{eqnarray}
and the third term gives an additional potential term.
The DM term is also  
known as an effect of spin-orbit coupling (SOC) in condensed matter physics.

We employ the background $SU(2)$ gauge field 
with
the field strength 
$F_{\mu\nu} = \partial_\mu A_\nu -  \partial_\nu A_\mu 
- i [A_\mu,A_\nu]$.
Here, we consider 
a nonzero
constant non-Abelian magnetic field 
\begin{eqnarray}
  F_{12} = \kappa^2 \sigma_3 ,
  \label{eq:bg-gauge}
\end{eqnarray}
with a constant 
$\kappa$. 
Because of this field strength, 
the flavor symmetry is explicitly broken 
to the $U(1)$ subgroup generated by 
$\sigma_3$.
The gauge potential leading 
to the field strength 
given in Eq.~(\ref{eq:bg-gauge}) is 
for instance 
\begin{equation}
    \begin{split}
        & A_{0}^a = (0,0,0) \\
        & A_{1}^a=-\kappa(\cos \vartheta,-\sin \vartheta, 0) \\
        & A_{2}^a=-\kappa(\sin \vartheta, \cos \vartheta, 0)
    \end{split}
    \label{eq:SOC1}
\end{equation}
or
 \begin{eqnarray}
    A_{\mu} &=& A_{\mu}^a \sigma_a 
    = - \kappa (0,
    \cos \vartheta \sigma_1 - \sin \vartheta \sigma_2,
    \sin \vartheta \sigma_1 + \cos \vartheta \sigma_2)
    \nonumber\\
    &=& - \kappa \left(0,
    \left(\begin{array}{cc}
                  0 & e^{i \vartheta}\\
    e^{-i \vartheta} & 0
    \end{array}\right),
    \left(\begin{array}{cc}
                  0 & -ie^{i \vartheta}\\
    ie^{-i \vartheta} & 0
    \end{array}\right)
    \right)
    \label{eq:SOC2}
 \end{eqnarray}
with a constant 
$\vartheta$.
This parameter $\vartheta$ corresponds to just a gauge choice. 
Nevertheless 
each $\vartheta$ gives 
a different looking term 
in the Lagrangian. 
 In particular, the case of 
 $\vartheta = 0$ is called  
 the Dresselhaus SOC 
 \begin{eqnarray}
    A_{\mu} = A_{\mu}^a \sigma_a 
    = - \kappa (0,\sigma_1, \sigma_2 ,0) 
    \label{eq:Dresselhaus-SOC}
 \end{eqnarray}
yielding the DM term in the form of 
	\begin{equation}
		{\cal H}_{\rm DM} = \kappa 
  \n\cdot (\nabla\times \n),
		\label{Dresselhaus-Bloch}
	\end{equation}
 and the case of 
 $\vartheta = -\pi/2$ is called 
 the Rashba SOC 
  \begin{eqnarray}
    A_\mu = A_\mu^a \sigma_a = \kappa (0,-\sigma_2, \sigma_1 ,0) 
 \end{eqnarray}
	that yields the DM term in the form of 
	\begin{equation}
		{\cal H}_{\rm DM}= \kappa 
  (\n\cdot \nabla n_3 - n_3\nabla\cdot \n). 
		\label{Rashba-Neel}
	\end{equation}
The DM terms in Eqs.~\eqref{Dresselhaus-Bloch} and \eqref{Rashba-Neel} 
are known to admit magnetic domain walls  of 
the Bloch and N\'eel types, 
respectively.
They also admit magnetic skyrmions 
of 
the Bloch and N\'eel types, 
respectively.
We emphasize that 
these two terms as well as 
the terms for general $\vartheta$ look different 
but physically are equivalent to 
each other under a field redefinition (gauge transformation)
because they give the same field strength 
in Eq.~(\ref{eq:bg-gauge}).

\begin{figure}[h]
    \centering
    \begin{tabular}{cc}
      \includegraphics[width=0.4\columnwidth]{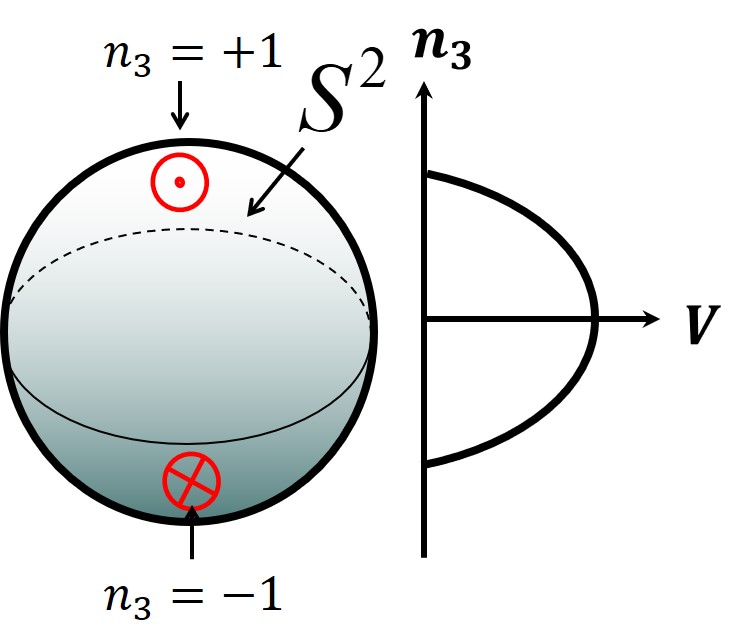}   & 
      \includegraphics[width=0.4\columnwidth]{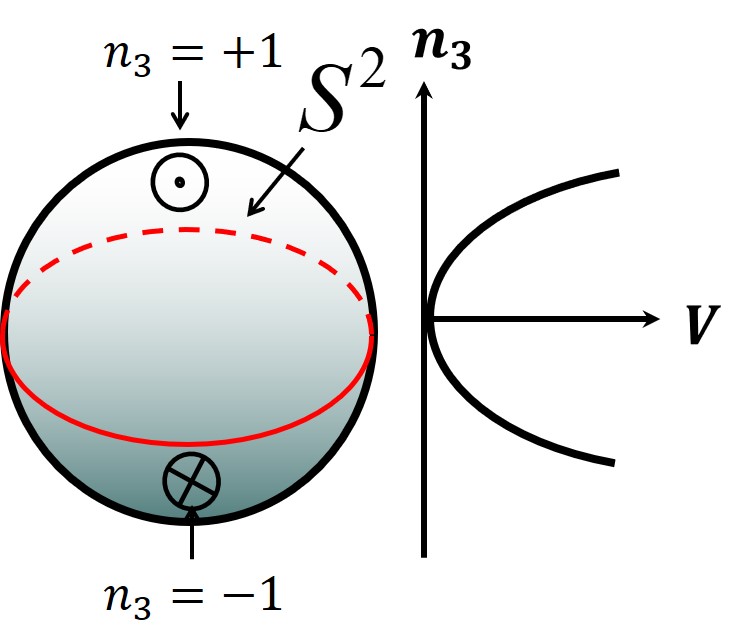}\\
       (a)  &  (b)
    \end{tabular}
    \caption{The potentials with vacua denoted in red.
    (a) Easy-axis potential. 
    Vacua are the north ($n_3=+1$) and south 
    ($n_3=-1$) poles.
    (b) Easy-plane potential. 
     Vacua are on the equator ($n_3=0$).}
    \label{fig:potential}
\end{figure}

The total potential term is\footnote{
Because of the additional constant term $\kappa^2$ in the potential, 
SUSY is broken if the original Lagrangian is SUSY.
} 
\begin{align}
V_{\rm tot} 
& = \frac{1}{2}\left(\A_{k} \times \n\right)^{2}+m^{2}(1- n_{3}^{2}) 
\notag\\
& =\frac{1}{2}\left|\A_{k}\right|^{2}|\n|^{2}-\frac{1}{2}\left(\A_{k} \cdot \n\right)^{2}+m^{2} (1-n_{3}^{2}) \notag\\
& =\kappa^{2}-\frac{\kappa^{2}}{2}\left(n_{1}^{2}+n_{2}^{2}\right)
+m^{2} (1-n_{3}^{2}) \notag\\
& = \left(\frac{\kappa^{2}}{2}-m^{2}\right) n_{3}^{2} 
+ 
\frac{\kappa^{2}}{2}+m^2.
\end{align}
Apart from the constant terms, this potential is called 
\begin{eqnarray}
\left\{
\begin{array}{cccl}
{\rm (i)}  &\kappa^2 - 2m^2>0   &:  &  \mbox{Easy-plane}\\
{\rm (ii)} &\kappa^2  -2m^2=0   &:   &  \mbox{No potential}\\
{\rm (iii)} &\kappa^2  -2m^2<0  &:    &  \mbox{Easy-axis}\\
\end{array}
\right. . \label{eq:potential}
\end{eqnarray}
These potentials are drawn schematically in Fig.~\ref{fig:potential}.

\begin{figure}[h]
    \centering
    \includegraphics[width=0.8\columnwidth]{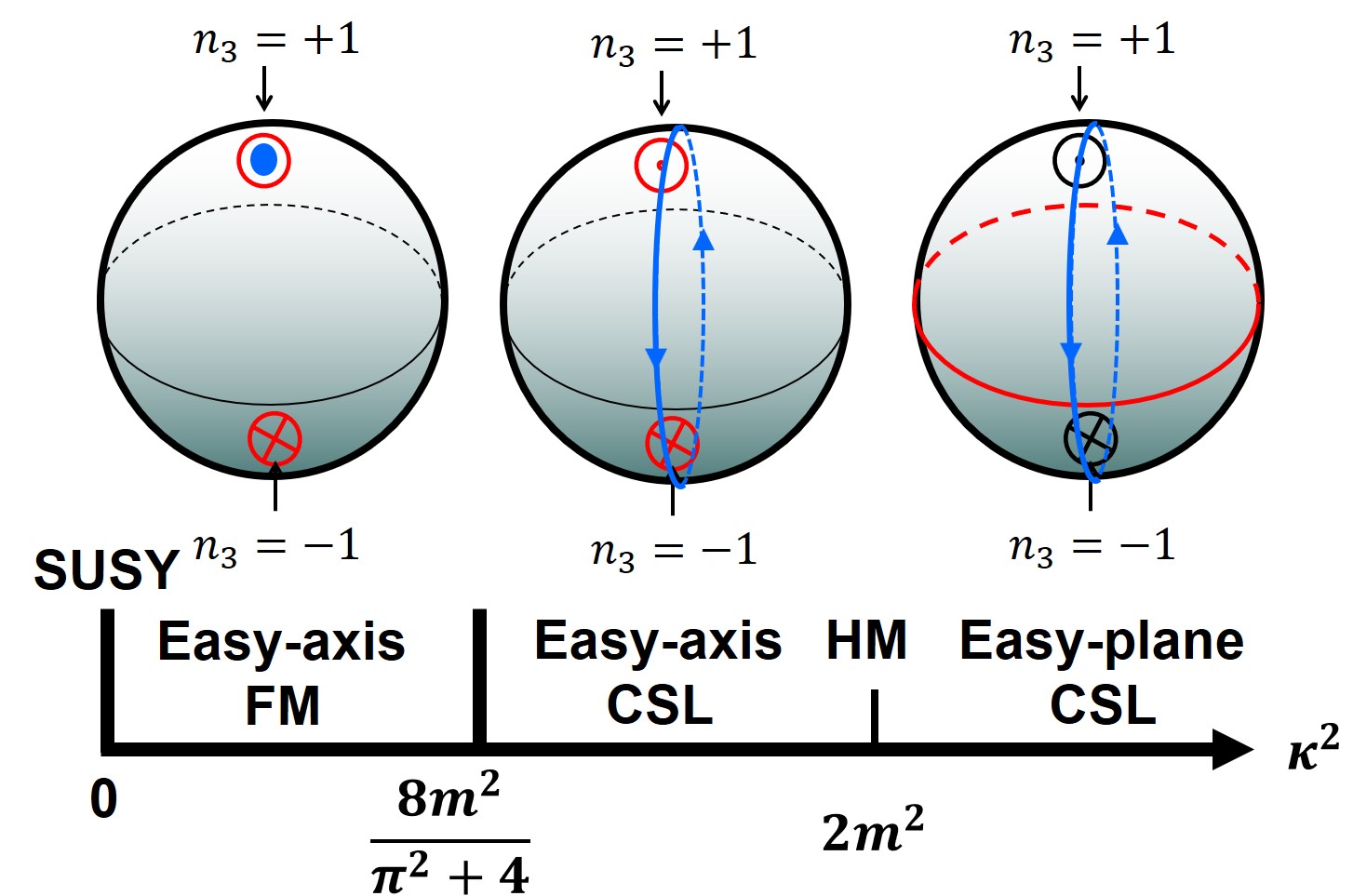}
    \caption{The Phase diagram of the chiral magnet from D-branes. 
    FM and HM denote ferromagnetic 
    and helimagnetic, 
    respectively.
    In the FM phase,
    the ground state denoted by a blue dot is either 
    the north pole $n_3=+1$ or south pole $n_3=-1$.
    In the easy-axis (plane) CSL, 
    while the vacua are the north and south poles (equator), 
    the ground state is a CSL represented by a blue circle. 
    }
    \label{fig:phase-diagram}
\end{figure}

As we show in a later section, 
the ground state is not a uniform state 
but is inhomogeneous, 
or forms a CSL, 
when the following inequality holds:
\begin{equation}
    4\left|\kappa^{2}-2 m^{2}\right|<\kappa^{2} \pi^{2} .
\end{equation}
Then, there are 
a ferromagnetic phase with the easy-axis potential, 
the CSL phases 
with the easy-axis or easy-plane potential, 
and a helimagnetic phase at the boundary between the latter. 
In summary, 
by gradually increasing the DM interaction $\kappa$,
there are the following phases:  
\begin{eqnarray}
\left\{
\begin{array}{ccl}
 \kappa^2 =0 & : & {\rm SUSY } \\   
 0 \leq \kappa^2 \leq \displaystyle{\frac{8m^2}{\pi^2+4}} &: 
 & {\rm Ferromagnetic}\\
\displaystyle{\frac{8m^2}{\pi^2+4}} \leq \kappa^2 
 < 2m^2 & :  &
  \mbox{Easy-axis CSL}\\
  \kappa^2 =2 m^2& :  
  & \mbox{Helimagnetic}\\
  2m^2 < \kappa^2 & :  &
  \mbox{Easy-plane CSL}
 \end{array}
 \right. .
\end{eqnarray}
The phase diagram is given in Fig.~\ref{fig:phase-diagram}.

\section{Chiral Magnets from D-branes}\label{sec:dbrane}

In this section, 
we introduce two D-brane configurations 
in type-IIA/B string theory.
In Sec.~\ref{subsec:dbrane1}, 
we give the Hanany-Witten brane configuration 
consisting of D3, D6, NS5-branes 
in type-IIB string theory.
In Sec.~\ref{subsec:dbrane2}, 
we give fractional D2-D6 branes on 
the Eguchi-Hanson manifold 
In both cases, 
the $O(3)$ nonlinear sigma model 
is realized on the color D-brane, 
and we further introduce 
non-Abelian magnetic fluxes on 
the flavor D-branes 
inducing the DM interaction.

\subsection{Hanany-Witten 
D-brane configuration}
\label{subsec:dbrane1}

As mentioned, the gauge theory introduced in the last section 
can be made ${\cal N}=2$ SUSY  
by introducing the Higgs scalar fields $\tilde\Phi$ 
and adding fermionic superpartners (Higgsinos) 
for hypermultiplets, 
and gauginos for gauge or vector multiplets  \cite{Eto:2006pg}.
Then, the theory can be realized by D-brane configurations.  
We first consider the Hanany-Witten brane configuration in type IIB string theory 
\cite{Hanany:1996ie,Giveon:1998sr}.
Here, we construct a more general Grassmann sigma model with 
the target space 
\begin{eqnarray}
    Gr_{\NF,\NC} = \frac{SU(\NF)}{SU(\NC) \times 
SU(\NF-\NC) \times U(1)}
\end{eqnarray}
by considering 
the ${\cal N}=2$ SUSY 
$U(\NC)$ gauge theory 
coupled with $\NF$ hypermultiplets 
in the fundamental representation, 
and later restrict ourselves to 
$\NF=2 $ and $\NC=1$ for 
the ${\mathbb C}P^1$ model.\footnote{
More precisely, for ${\mathcal N}=2$ SUSY the vacuum manifolds are 
cotangent bundles $T^*Gr_{\NF,\NC}$,  $T^* {\mathbb C}P^1$ 
and so on.
}

\begin{figure}[h]
 \begin{center}
 \begin{tabular}{cc}
 \includegraphics[width= 70mm]{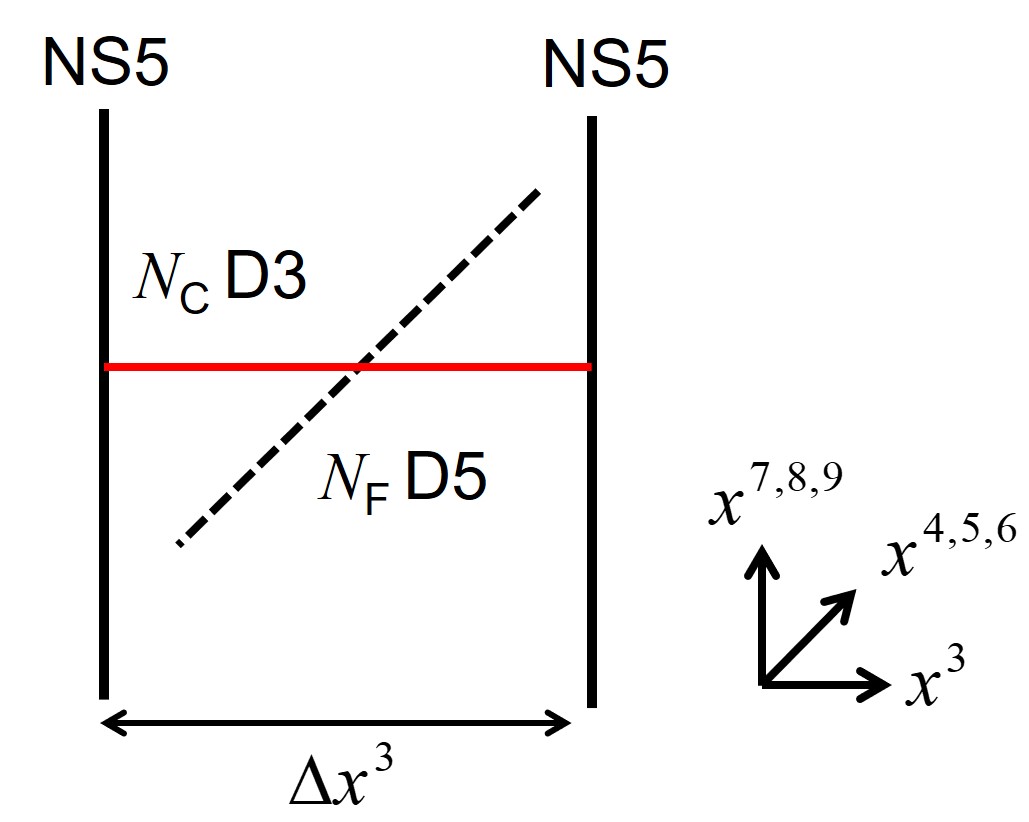}
 &\includegraphics[width= 70mm]{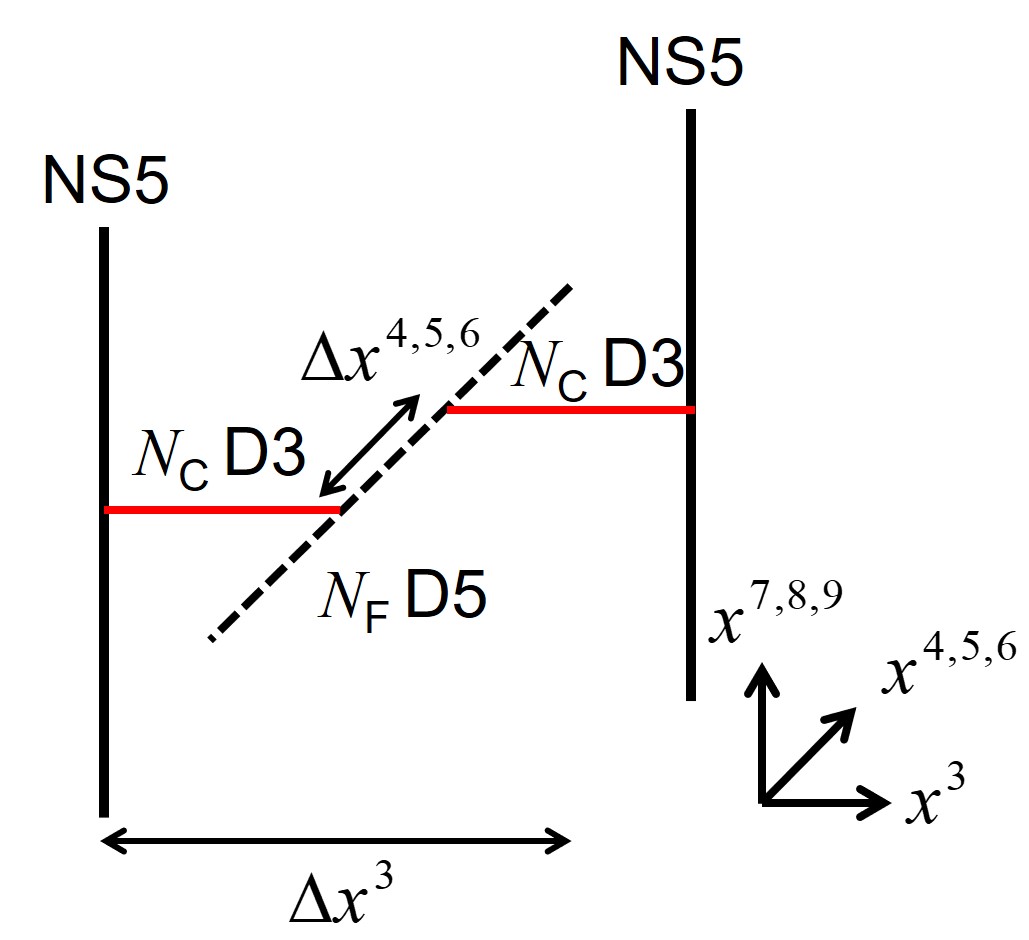} \\
  a) & b) \\
 \includegraphics[width= 70mm]{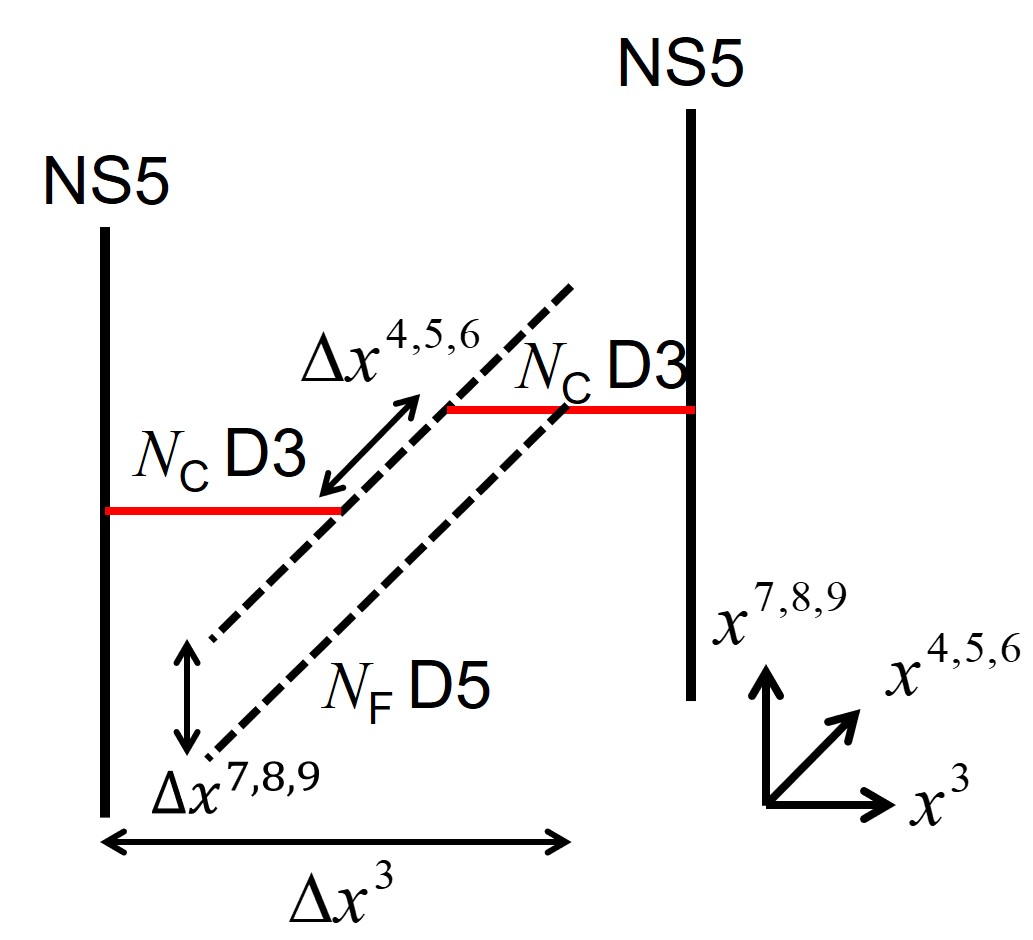} & 
 \includegraphics[width= 70mm]{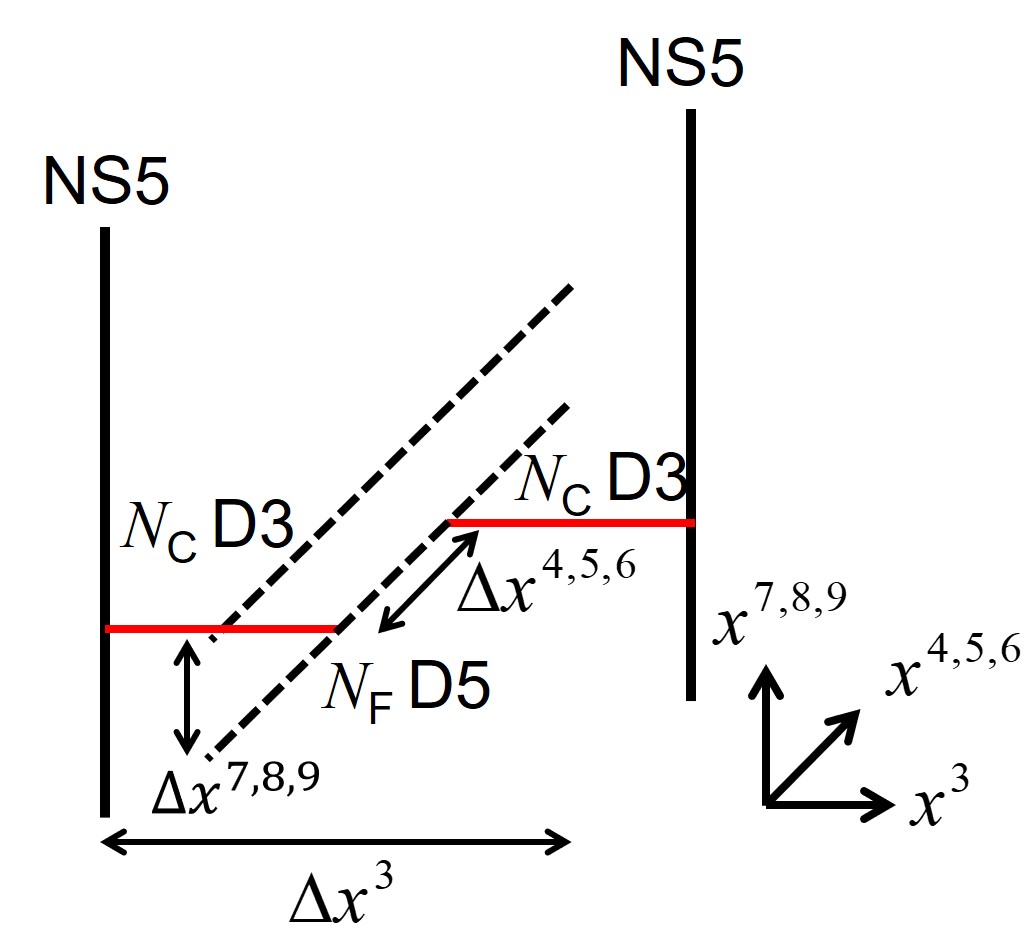}
 \\
 c) & d)
 \end{tabular}
 \end{center}
 \caption{The Hanany-Witten brane configuration.
 The $U(\NC)$ gauge theory is realized 
on the worldvolume of the $\NC$ D3-branes. 
Strings connecting two D3-branes give gauge multiplets 
while string connecting a D3-brane and a D6-brane 
give hypermultiplets.
The separation 
$\Delta x^3$ of the two NS5-branes into 
the $x^3$ direction corresponds to $1/g^2$. 
a) Hypermultiplets are massless, 
and do not have a VEV.
b) A triplet of FI-terms is introduced 
by separation 
$(\Delta x^4,\Delta x^5,\Delta x^6)$ of 
the two NS5-branes  into the $x^{4,5,6}$ directions.
c) and d) The masses of the hypermultiplets 
are introduced by the separation of D5-branes
 into the $x^{7,8,9}$ directions.
}
 \label{fig:HW-brane}
\end{figure}

In Table \ref{tab:HW-brane}, we summarize 
the directions in which the D-branes extend, 
and 
 the brane configuration is 
schematically drawn 
in Fig.~\ref{fig:HW-brane}.
\begin{table}[h]
\begin{center}
\begin{tabular}{c|cccccccccc}
\hline
&$x^0$&$x^1$&$x^2$&$x^3$&$x^4$&$x^5$&$x^6$&$x^7$&$x^8$&$x^9$\\
\hline
$N_{\rm C}$ $\rm D3$&$\circ$&$\circ$&$\circ$&$\circ$&$-$&$-$&$-$&$-$&$-$&$-$ \\
\hline
$N_{\rm F}$ $\rm D5$&$\circ$  &
$\circ*$ &$\circ*$&$-$&$\circ$&$\circ$&$\circ$&$-$&$-$&$-$ \\
\hline 
$2\; {\rm NS}5$&$\circ$&$\circ$&$\circ$&$-$&$-$&$-$&$-$&$\circ$&$\circ$&$\circ$ \\
\hline
\end{tabular}
\end{center}
\caption{The Hanany-Witten brane configuration.  
Branes are extended along directions denoted by $\circ$, and are not extended
along directions denoted by $-$. 
$*$ denotes a background gauge field strength 
making D5-branes magnetized.
}
\label{tab:HW-brane}
\end{table}
In Fig.~\ref{fig:HW-brane} a),
$\NC$ D3-branes are stretched between 
two ${\rm NS}5$ branes 
separated into the $x^3$ direction.
The $U(\NC)$ gauge theory is realized on the $\NC$ 
coincident D3-brane world-volume. 
The $\rm D3$-brane world-volume 
have the finite length $\Delta x^3$ between 
two ${\rm NS}5$-branes,  and therefore 
the $\rm D3$-brane world-volume theory is
$(2+1)$-dimensional 
$U(\NC)$ gauge theory.\footnote{
The gauge coupling is given by 
$\frac{1}{g^2} = |\Delta x^3| \tau_3 l_s^4 
= \frac{|\Delta x^3|}{g_s}$, with  
the string coupling constant $g_s$ 
and string length $l_s$ 
in type IIB string theory
and the D3-brane tension  $\tau_3 = 1/g_s l_s^4$.
}

The positions of the $\NF$ $\rm D5$ branes 
in the $x^7$-, $x^8$- and $x^9$-directions coincide 
with those of the $\rm D3$ branes. 
Strings which connect between D3 and D5 branes give rise to 
the $\NF$ hypermultiplets (the Higgs fields $\Phi,\tilde \Phi$ and Higgsinos) 
in the D3-brane worldvolume theory.

Next, we put the system into the Higgs phase by separating
the positions of the two NS5-branes 
in the $x^{4,5,6}$ directions,
$(\Delta x^4,\Delta x^5,\Delta x^6) \neq 0$, 
as in Fig.~\ref{fig:HW-brane} b). 
This gives rise to the triplet of 
the Fayet-Iliopoulos(FI) parameters $c^a$ 
\cite{Hanany:2003hp,Hanany:2004ea}, 
and we choose it as $c^a = (0,0,v^2=\Delta x^4/g_s l_s^2 >0)$.
Then, the D3-brane worldvolue is cut 
and each segment of a D3-brane ends on one D5-brane.

In the third step, 
we introduce masses to the hypermultiplets, 
by separating 
the positions of the D5-branes 
into the $x^{7,8,9}$ directions 
as in  Fig.~\ref{fig:HW-brane} c) and 
d). 
This gives rise to triplet 
masses to the hypermultiplets. 
We consider real masses with 
$\Delta x^7=0$ for simplicity.

The vacua of the D3-brane 
worldvolume theory can be considered as follows.
As shown in Fig.~\ref{fig:HW-brane} c) 
and d), each $\rm D3$ 
brane ends on one of the $\rm D5$ branes, 
on each of which 
at most one $\rm D3$ brane can end,  
which is known as the s-rule~\cite{Hanany:1996ie}. 
There are ${}_{\NF}C_{\NC}=\NF !/\NC !(\NF-\NC) !$ vacua in 
the Grassmann sigma model 
\cite{Arai:2003tc}.
The case of $\NF=2, \NC=1$ that we concern in this paper, 
there are two vacua 
as in Fig.~\ref{fig:HW-brane} c) 
and d), 
corresponding to 
the 
antipodal 
points on ${\mathbb C}P^1 \simeq S^2$.

Finally, we turn on 
a background gauge field 
in Eq.~(\ref{eq:bg-gauge}), 
Eq.~(\ref{eq:SOC1}) or
(\ref{eq:SOC2}),
on the $x^1$-$x^2$ plane 
in the D5-brane worldvolume, 
which are the common directions 
with D3-brane worldvolume 
not shown in Fig.~\ref{fig:HW-brane}.
Such a background gauge field 
makes D5-branes magnetized \cite{Bachas:1995ik,Berkooz:1996km,Blumenhagen:2000wh,Angelantonj:2000hi,
Cremades:2004wa,Kikuchi:2023awm,Abe:2021uxb}, 
and 
the $SU(\NF=2)$ symmetry, 
which is a gauge symmetry on 
the D5-banes, 
is 
spontaneously broken 
to the $U(1)$ subgroup generated by 
$\sigma_3$.
This background gauge field 
induces the DM term on 
the D3 brane worldvolume theory, 
as in the second term in 
Eq.~(\ref{eq:generalized-DM}), 
or more explicitly 
Eq.~(\ref{Dresselhaus-Bloch})
or (\ref{Rashba-Neel}). 
SUSY is completely broken at this step.
We will see that 
this final step gives very nontrivial physics. 
In particular,  we will find that 
the ground states 
are not uniform anymore in general
as summarized in Fig.~\ref{fig:phase-diagram}.
Before discussing that, we give another useful 
brane configuration related to this brane configuration.

\subsection{D2-D6 system on Eguchi-Hanson space}
\label{subsec:dbrane2}

We take a T-duality in the $x^3$ direction 
along which the NS5-branes are separated 
to obtain 
a D$2$-D$6$ system with $\NC$ D$2$-branes 
and $\NF$ D$6$-branes in type-IIA string theory 
\cite{Eto:2004vy}, 
see Table \ref{tab:brane-ALE}.
\begin{table}[h]
\begin{center}
\begin{tabular}{c|cccccccccc}
\hline
&$x^0$&$x^1$&$x^2$&$x^3$&$x^4$&$x^5$&$x^6$&$x^7$&$x^8$&$x^9$\\
\hline
$N_{\rm C}$ $\rm D2$&$\circ$&$\circ$&$\circ$&$-$&$-$&$-$&$-$&$-$&$-$&$-$ \\
\hline
$N_{\rm F}$ $\rm D6$&$\circ$&$\circ*$ &$\circ*$ &$\circ$&$\circ$&$\circ$&$\circ$&$-$&$-$&$-$ \\
\hline 
{\rm ALE} ${\mathbb C}^2/{\mathbb Z}_2$ &$-$&$-$&$-$&$\circ$&$\circ$&$\circ$&$\circ$&$-$&$-$&$-$ \\
\hline
\end{tabular}
\end{center}
\caption{D2-D6-brane configuration 
on the Eguchi-Hanson manifold 
in type-IIA string theory.
Branes are extended along directions denoted by $\circ$, and are not extended
along directions denoted by $-$. 
}
\label{tab:brane-ALE}
\end{table}

\begin{figure}[thb]
\begin{center}
\begin{tabular}{ccc}
  \includegraphics[width=4.7cm,clip]{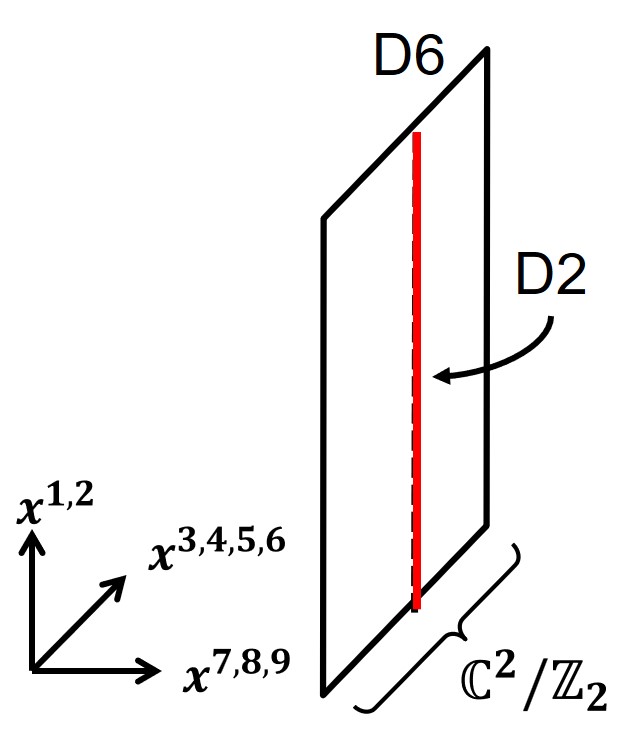}
 &
   \includegraphics[width=5cm,clip]{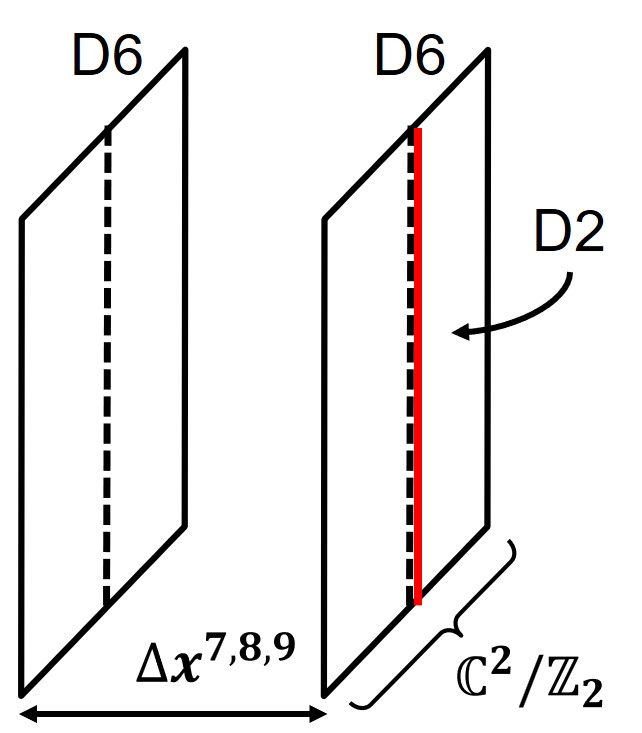}
   & 
   \includegraphics[width=5cm,clip]{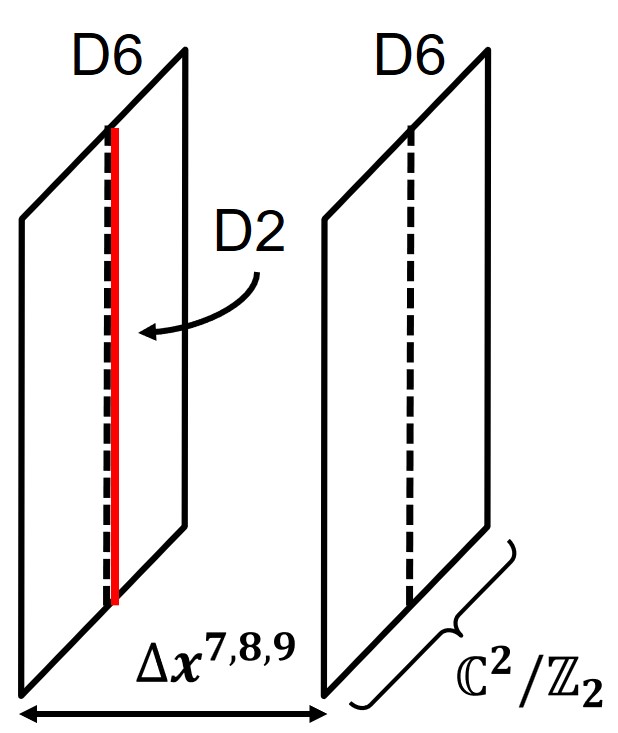}
\\
a) &    b)  &  c) 
\end{tabular}
\end{center}
\caption{
The D2-D6 branes in the Eguchi-Hanson manifold in type-IIA string theory. 
The case of $\NF=2$ and $\NC=1$ is drawn. 
The dashed lines denote 
the fixed points (orbifold singularity) of the ${\mathbb Z}_2$ action 
on the orbifold ${\mathbb C}^2/{\mathbb Z}_2$. 
The singularity is blown up by $S^2$ to become 
the Eguchi-Hanson manifold. 
The D2-branes are fractional D2-branes, 
that is D4-branes two of whose worldvolume wrap 
$S^2$.
a)
The D-brane configuration for 
massless hypermultiplets in the fundamental representation.
b) The brane configuration for 
massive hypermultiplets.
Hypermultiplets coming from 
strings stretched between D$2$- and D$6$-branes 
become massive by placing D$6$-branes with distances 
in the $x^{7,8,9}$-coordinates.
}
\label{fig:brane-ALE}
\end{figure}

The hypermultiplets come from 
strings stretched between D$2$- and D$6$-branes.
First, we consider the case that 
all hypermultiplets are massless. 
In this duality, NS5-branes are mapped to 
a hyper-K\"ahler geometry.
The orthogonal space ${\mathbb C}^2$ 
(the $x^{3,4,5,6}$ directions)
perpendicular to the D$2$-branes 
inside the D$6$-brane world-volume 
is divided by ${\mathbb Z}_2$,
and there is a constant self-dual NS-NS $B$-field 
on ${\mathbb C}^2/{\mathbb Z}_2$. 
The asymptotically locally Euclidean space (ALE) space of the $A_1$-type, the Eguchi-Hanson space $T^* {\mathbb C}P^1$, 
is obtained by blowing up the orbifold singularity by 
inserting $S^2$.\footnote{
The FI-parameter $c$  
blows up the ${\mathbb Z}_2$ orbifold singularity 
with replacing it by $S^2$ of the area 
 $A = c g_s l_s^{p+1} = c / \tau_p$ 
 for the D$p$-D$(p+4)$ system 
 (in our case $p=2$). 
 Here, 
 $g_s$ is the string coupling constant  and 
 $l_s = \sqrt {\alpha'}$ is the string length 
 in type-IIA string theory.
  Our D$p(=2)$ brane is fractional D$p$-brane, 
 D$(p+2)(=4)$ brane wrapping around $S^2$.
The gauge coupling constant $g$ is given by 
$\frac{1}{g^2} = b \, \tau_{p+2} \, l_s^2 
   = {b \over \, g_s l_s^{p-3} }$
with  the D($p+2$)-brane tension 
$\tau_{p+2} = 1/ g_s l_s^{p+3}$
and 
 the $B$-field flux 
 $b \sim A B_{ij}$
 integrated over the $S^2$.
The gauge theory limit 
(with gravity and higher derivative corrections decoupled) 
is taken as 
$l_s \to 0$ with keeping $g^2$ and $c$ fixed. 
} 
The D$2$-branes are actually fractional D$2$-branes, 
that is, D$4$-branes two of whose spatial directions 
in the whole 4+1 dimensional world-volumes
are wrapped around $S^2$, which blows up 
the orbifold singularity of ${\mathbb C}^2/{\mathbb Z}_2$.
 Thus, the positions of fractional  D$2$-branes 
inside the D$6$-branes  
are fixed at the fixed point of the 
${\mathbb Z}_2$ action, 
and thus there are no adjoint hypermultiplets
on the D$2$-brane world-volume theory, 
see Fig.~\ref{fig:brane-ALE} a).

The $\NC$ D$2$-branes can be interpreted as 
Yang-Mills instantons (with the instanton number $\NC$) on the Eguchi-Hanson manifold 
in the effective $U(\NF)$ gauge theory on 
the D$6$-branes. 
The Kronheimer-Nakajima construction
\cite{Kronheimer1990}
 of the moduli space of instantons 
in the ALE space gives the same moduli space 
of vacua $T^* Gr_{\NF,\NC}$ 
as that of the Hanany-Witten 
brane configuration.

Now we turn on the masses of the hypermultiplets.
The masses of the hypermultiplets correspond to the positions of 
the D$6$-branes in the $x^{7,8,9}$-directions.  
Thus, 
the hypermultiplets coming from 
strings stretched between D$2$- and D$6$-branes 
become massive by this separation. 
We assume real masses for hypermultiplets 
by placing D6-branes parallel along the $x^7$-direction. 
To be consistent with the s-rule in the T-dual picture 
at most one D2-brane can be absorbed into 
the ${\mathbb Z}_2$ fixed point of one D6-brane.

Again, we finally turn on 
a background gauge field 
(\ref{eq:bg-gauge}) 
making D6-branes magnetized.
This yields the DM term, given in 
Eq.~(\ref{eq:SOC1}) or
(\ref{eq:SOC2}), 
on the D2-brane worldvolume theory.

In the following sections, 
we discuss brane configurations 
corresponding to topological solitons 
and modulated ground states. 
For such purpose, we will see 
that the latter brane configuration is 
simpler.

\section{Domain Walls and Chiral Soliton Lattice Phases as D-branes
} \label{sec:wall-CSL}

In this section, 
we analytically consider topological solitons 
of codimension one 
and the ground states which are also codimension one (or uniform).
In Subsec.~\ref{sec:chiral-SG}, 
we reduce the model 
to the so-called chiral sine-Gordon model 
by assuming one dimensional dependence.
In Subsec.~\ref{sec:domain-wall}, 
we construct a magnetic domain wall 
as an excited state
in the ferromagnetic phase 
and a kinky D-brane configuration.
In Subsec.~\ref{sec:CSL-EA}, 
we construct the CSL phase with the easy-axis potential
and snaky D-brane configuration.
In Subsec.~\ref{sec:helimagnetic}, 
we study the helimagnetic phase.
In Subsec.~\ref{sec:CSL-EP}.
we construct the CSL phase with the easy-plane potential
and zigzag D-brane configuration.

\subsection{Chiral sine-Gordon model}\label{sec:chiral-SG}

First, 
we introduce rotated coordinates 
$(\tilde{x}^1,\tilde{x}^2)$ by 
\begin{equation}
    \left\{\begin{array}{l}
\tilde{x}^{1}=\cos \beta x^{1}+\sin \beta x^{2} \\
\tilde{x}^{2}=-\sin \beta x^{1}+\cos \beta x^{2} 
\end{array}\right. .
\label{eq:rotation}
\end{equation}
In the next subsection, we will take  $\tilde{x}^1$ as the direction 
perpendicular to 
the domain wall, 
$\tilde{x}^2$ the direction along 
the domain-wall worldvolume.
Writing $\frac{\partial}{\partial \tilde{x}^{k}}=\tilde{\partial}_{k}$, we have
\begin{equation}
  {\cal H}_{\rm DM} =  \A_{k} \cdot\left(\n \times \partial_{k} \n\right)=\tilde{\A}_{k} \cdot\left(\n \times \tilde{\partial}_{k}\n\right)
\end{equation}
where
\begin{equation}
    \begin{split}
        & \tilde{\A}_{1}=-\kappa(\cos \tilde{\vartheta},-\sin \tilde{\vartheta}, 0) \\
& \tilde{\A}_{2}=-\kappa(\sin \tilde{\vartheta}, \cos \tilde{\vartheta}, 0)
    \end{split}
\end{equation}
with $\tilde{\vartheta} \equiv \vartheta-\beta$.

We employ the ansatz 
for configurations depending on 
only one direction that we take $\tilde{x}^1$
(domain walls and
chiral soliton lattices) 
of the form
\begin{equation}
    \n=\left(\cos \phi \sin f(\tilde{x}^{1}), \sin \phi \sin f(\tilde{x}^{1}), \cos f(\tilde{x}^{1})\right).
    \label{eq:wall-ansatz}
\end{equation}
Then, the DM interaction can be written as
\begin{align}
& 
  {\cal H}_{\rm DM} = 
\tilde{\A}_{k} \cdot\left(\n \times \tilde{\partial}_{k}\n\right)=\tilde{\A}_{1} \cdot\left(\n \times \tilde{\partial}_{1}\n\right)  
=\kappa \sin (\tilde{\vartheta} + \phi) \tilde{\partial}_{1} f
\end{align}
So, one gets the chiral sine-Gordon model
\begin{align}
{\cal H}_\text{SG}=\frac{1}{2}\left(\tilde{\partial}_{1} f\right)^{2}
+\kappa \sin (\tilde{\vartheta}  + \phi) \tilde{\partial}_{1} f +\left(\frac{\kappa^{2}}{2}-m^{2}\right) \cos ^{2} f +\frac{\kappa^{2}}{2} + m^2.
\label{eq:chiral-sine-Gordon_Hamiltonian_easy-plane}
\end{align}
The second term 
is a total derivative term 
specific for the {\it chiral} sine-Gordon model, 
or a topological term counting 
the number of sine-Gordon solitons. 
Note that this term does not contribute to the equation of motion.

The trivial vacuum solutions are given by
\begin{equation}
    f_\text{vac} = \left\{
    \begin{array}{clll}
       \left(l + \frac{1}{2}\right)\pi & ~~\text{if} &
       \kappa^2-2m^2 > 0 &~\text{(easy-plane)}
        \\
       l\pi & ~~\text{if} & \kappa^2-2m^2 < 0 &~\text{(easy-axis)}
    \end{array}
    \right.
\end{equation}
with an integer $l$. When the potential term vanishes, i.e., $\kappa^2 -2m^2=0$, any constant can be the vacuum solution.

The chiral sine-Gordon model 
or chiral double sine-Gordon model also appears in 
QCD at finite density in the presence 
of strong magnetic field
\cite{Son:2007ny,Brauner:2016pko}  
or rapid rotation 
\cite{Nishimura:2020odq,Eto:2021gyy}.
In such cases, 
the Wess-Zumino-Witten term 
\cite{Son:2004tq}
gives a topological term 
instead of the DM term.

\subsection{Domain walls in ferromagnetic phase with easy-axis potential}\label{sec:domain-wall}

First, we consider the ferromagnetic phase. 
In this phase, there are two vacua 
corresponding to the north and south poles $n_3 =\pm 1$. 
In terms of the D-brane configurations, they correspond 
to the straight D-branes 
in Fig.~\ref{fig:HW-brane} c) and d) 
for the Hanany-Witten brane configuration in type-IIB string theory 
and Fig.~\ref{fig:brane-ALE} b) and c) for the D2-D6-ALE system 
in type-IIA string theory.

\begin{figure}[h]
\begin{center}
\begin{tabular}{cc}
\includegraphics[width=3cm,clip]{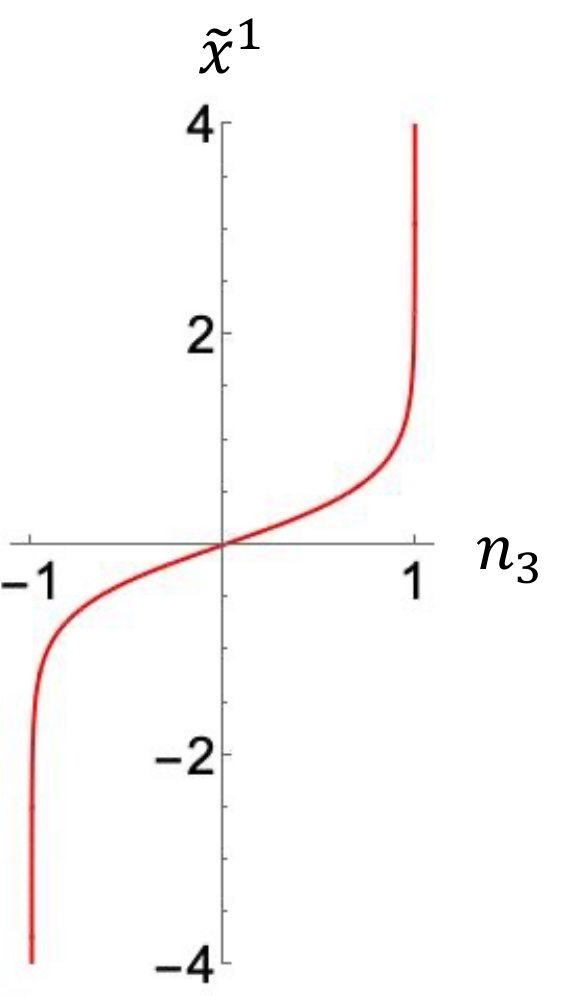}
&
\includegraphics[width=6.5cm,clip]{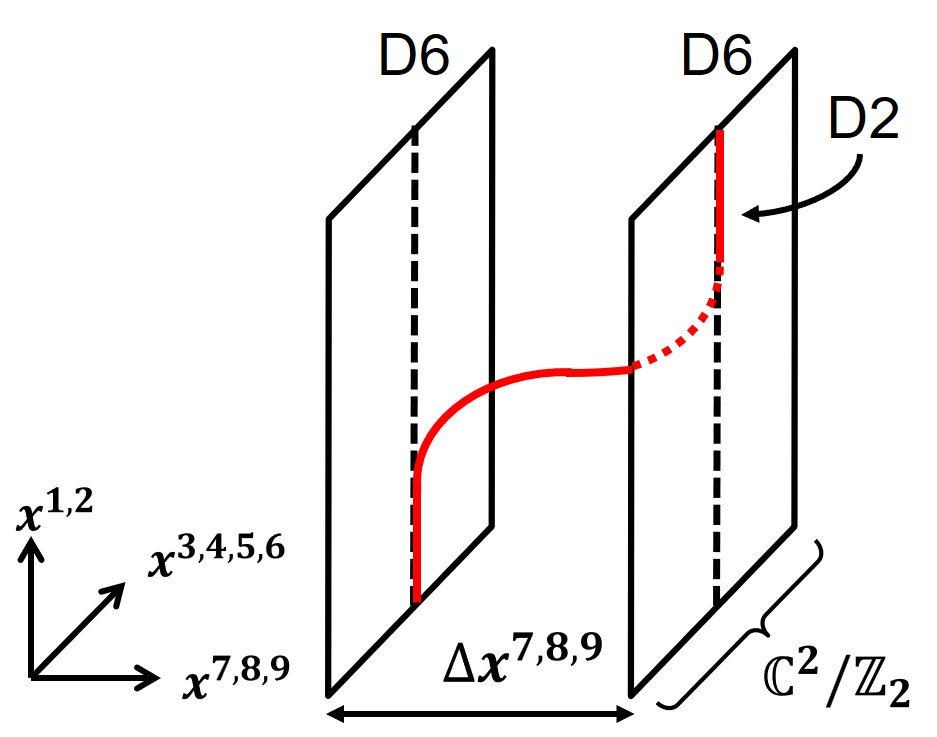}
\includegraphics[width=4.5cm,clip]{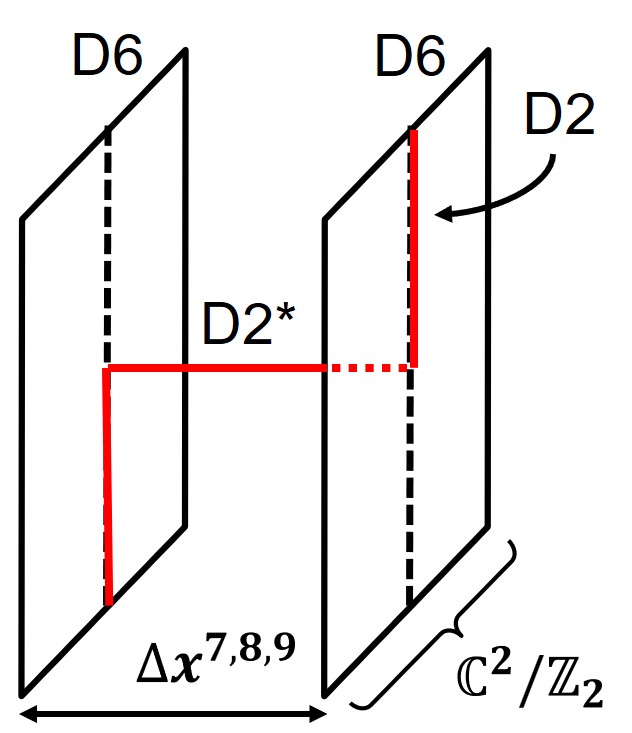}
\\
(a) & (b)
\end{tabular}

\includegraphics[width=140mm]{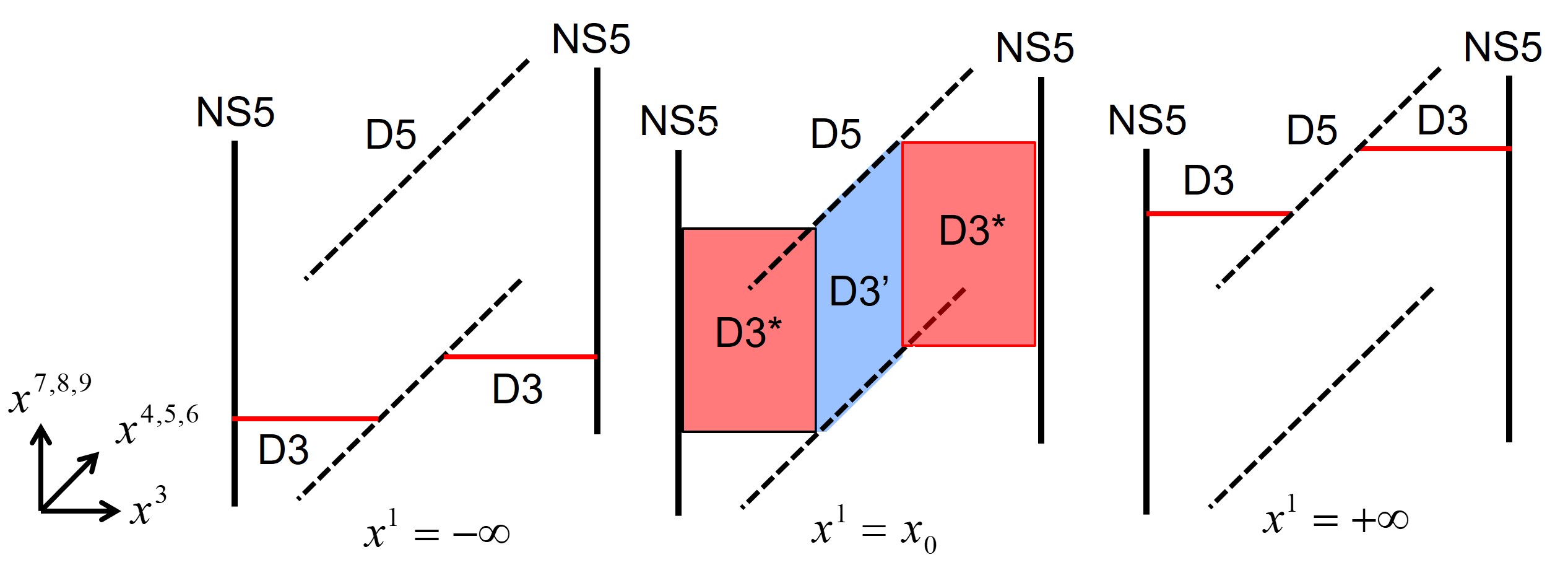}\\
(c)\\
\includegraphics[width=60mm]{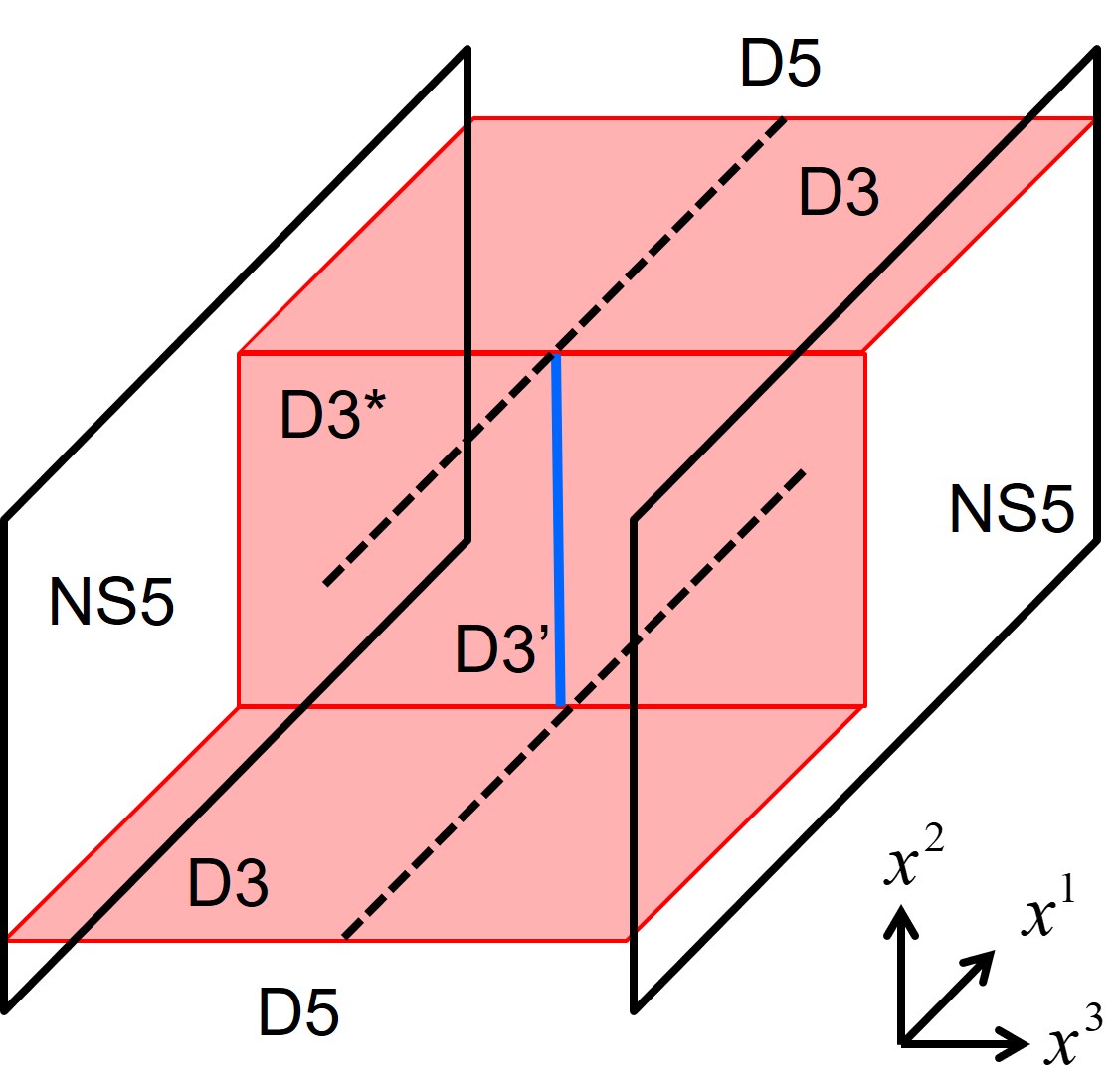}
\\
(d) 
\end{center}
\caption{(a) A single magnetic domain wall 
with the easy potential at the critical point ($\kappa=1, m^2=(\pi^2 - 4)/8 $). 
 The corresponding kinky 
brane configuration in 
(b) the D2-D5-ALE system
and 
(c), (d) the Hanany-Witten brane configuration.
\label{fig:easy-axis-wall}
}
\end{figure}

Then, we consider a magnetic domain wall interpolating these two vacua.
In this case, the energy per unit length in the $\tilde{x}^2$-direction is given by
\begin{align}
    E[f]&\equiv\int d\tilde{x}^1{\cal H}_\text{SG}
    \notag\\
    &=\int d\tilde{x}^1 \left[\frac{1}{2}\left(\tilde{\partial}_{1} f\right)^{2}+\kappa \sin (\tilde{\vartheta} + \phi) \tilde{\partial}_{1} f +\left(m^{2}-\frac{\kappa^{2}}{2}\right) \sin ^{2} f \right] + E_\text{vac}^\text{e.a.} \ ,
    \label{eq:chiral-sG_energy_easy-axis}
\end{align}
where $E_\text{vac}^\text{e.a.}$ denotes the vacuum energy with the easy-axis potential.
Let us study a single kink solution.
The (anti-)BPS equation for the magnetic domain wall can be given by
\begin{equation}
    \tilde{\partial}_{1} f=\pm\sqrt{2 m^{2}-\kappa^{2}} \sin f.
    \label{eq:BPS-eq-kink}
\end{equation}
The solution can be obtained as
\begin{equation}
    f_\text{kink}=2 \arctan \left[\exp \left(\pm\sqrt{2 m^{2}-\kappa^{2}} \tilde{x}^{1} + X\right)\right] \ ,
    \label{eq:kink_easy-axis}
\end{equation} 
where $X$ is a position moduli parameter.
If we choose the plus (minus) sign
for the (anti-)BPS soliton, the function $f$ monotonically increases (decreases).
The equation for the phase $\phi$ is simply given by
\begin{equation}
    \cos (\tilde{\vartheta} + \phi)=0 \ . 
    \label{eq:U(1)modulus}
\end{equation}
For the lowest energy kink solutions, the second term in the energy \eqref{eq:chiral-sG_energy_easy-axis}, which stems from the DM interaction, should be negative, so that $\phi$ is chosen as 
\begin{equation}
    \phi = \left\{ \begin{matrix}
        - \tilde{\vartheta} - \frac{\pi}{2} && \text{if~~} \kappa~\tilde{\pd}_1 f >0 \ ,
        \\
        - \tilde{\vartheta} + \frac{\pi}{2} && \text{if~~} \kappa~\tilde{\pd}_1 f <0 \ .
    \end{matrix}
    \right.
    \label{eq:phase_GS_easy-axis}
\end{equation}
For the phase $\phi$ giving the lowest energy kink solution, one finds that the energy difference between the single kink solution and the vacuum state is given by 
\begin{equation}
    E\left[f_{\text {kink }}\right] - E_\text{vac}^\text{e.a.}
    =2 \sqrt{2 m^{2}-\kappa^{2}} - |\kappa| \pi \ . 
    \label{eq:Energy-diff_kink-vac_easy-axis}
\end{equation}
See Fig. \ref{fig:easy-axis-wall} a) for a plot of the single domain wall  
(\ref{eq:kink_easy-axis}).
One of the most important role 
of the presence of the DM term is that 
this domain wall does not carry a 
$U(1)$ modulus \cite{Ross:2022vsa};
the $U(1)$ phase $\phi$ is fixed to be 
a constant determined from $\vartheta$
through Eq.~(\ref{eq:U(1)modulus}). 
When $\bf n$ at the domain wall 
($n_3=0$)
is parallel or orthogonal 
to the domain wall worldvolume, 
the wall is called Bloch or N\'eel, respectively, 
see Table~\ref{table:Bloch-Neel}.
\begin{table}
\begin{center}
\begin{tabular}{c|cccc}
     Type  & $\tilde \vartheta$ & $\phi$ & $S^1$ & to DW\\ \hline
    Bloch  & $0$ & $\pm \pi/2$  & $n_1=0$ & parallel \\  
    N\'eel & $\pm \pi/2$     & $\pm \pi$ & $n_2=0$ & orthogonal
\end{tabular}
\caption{Bloch or N\'eel type magnetic domain wall.
The parameter $\tilde \vartheta$,
$U(1)$ phase $\phi$ 
of the domain wall, 
$S^1$ submanifold 
inside target space $S^2$, 
and relation of $\n$ to the 
domain wall worldvolume.
}\label{table:Bloch-Neel}
\end{center}
\end{table}
\footnote{
This situation is 
in contrast to the case in the absence 
of the DM term in which 
a domain wall carries a 
$U(1)$ modulus $\phi$ 
\cite{Abraham:1992vb,Arai:2002xa,Arai:2003es}.
\label{footnote:U(1)}
}
We can change the direction of the domain wall worldvolume 
by changing $\beta$
in Eq.~(\ref{eq:rotation}) 
with the same ansatz in Eq.~(\ref{eq:wall-ansatz}).
Then, the $U(1)$ phase $\phi$ 
changes accordingly through 
Eq.~(\ref{eq:U(1)modulus}), 
and the angle between the 
$U(1)$ phase $\phi$ and 
the spatial direction of 
the domain wall worldvolume is 
preserved under a rotation.
This is well known and 
was reconfirmed in the effective 
theory of the domain wall 
\cite{Ross:2022vsa}.

Now let us discuss D-brane configurations for the magnetic domain wall.
In the D2-D6-ALE system 
in type-IIA string theory, 
the effective theory on 
the  D$2$-brane 
is the SUSY $U(\NC)$ gauge theory 
with massive $\NF$ hypermultiplets 
and the FI-term. 
For $\NC=1, \NF=2$ that we are considering, 
$\Sigma=m n_3$ 
for a single magnetic domain-wall solution 
is plotted as a function of $x^1$: 
$x^7 = \Sigma  (x^1)$ 
in Fig.~\ref{fig:easy-axis-wall} (b) left.\footnote{In the $U(\NC)$ case, the adjoint scalar field $\Sigma$ represents 
transverse fluctuations of the 
$\NC$ D$2$-branes.
In this case, 
the diagonal components of (the vacuum expectation value of) 
$\Sigma$, in the gauge of diagonal $\Sigma$,
can be identified as the positions of $\NC$ D$2$-branes 
along the $x^1$-coordinate.  
In that gauge, the $\NC$ diagonal components of $\Sigma$ represent 
for domain wall solutions. 
}
It represents a kinky D$2$-brane curved  
in the ($x^1$,$x^7$)-plane and  
the curve is determined  
by the solution
(\ref{eq:kink_easy-axis}) 
\cite{Lambert:1999ix,Eto:2004vy,Eto:2006mz,Eto:2007aw,Misumi:2014bsa}.
In the limit of a thin domain wall \cite{Hanany:2005bq}, 
the part of the kinky D2-brane can be 
regarded as a D2-brane extending into the $x^7$-direction 
instead of 
the $x^1$ direction (the codimension 
of the wall). We denote it by D2$^*$ (see Fig.~\ref{fig:easy-axis-wall}(b) right)
and the brane configuration 
is summarized in 
Table \ref{tab:brane-ALE-wall}.
\begin{table}[h]
\begin{center}
\begin{tabular}{c|cccccccccc}
\hline
&$x^0$&$x^1$&$x^2$&$x^3$&$x^4$&$x^5$&$x^6$&$x^7$&$x^8$&$x^9$\\
\hline
$N_{\rm C}$ $\rm D2$&$\circ$&$\circ$&$\circ$&$-$&$-$&$-$&$-$&$-$&$-$&$-$ \\
\hline
 $\rm D2^*$&$\circ$&$-$&$\circ$&$-$&$-$&$-$&$-$&$\circ$&$-$&$-$ \\
\hline
$N_{\rm F}$ $\rm D6$&$\circ$&$\circ*$ &$\circ*$ &$\circ$&$\circ$&$\circ$&$\circ$&$-$&$-$&$-$ \\
\hline 
{\rm ALE} ${\mathbb C}^2/{\mathbb Z}_2$ &$-$&$-$&$-$&$\circ$&$\circ$&$\circ$&$\circ$&$-$&$-$&$-$ \\
\hline
\end{tabular}
\end{center}
\caption{The domain wall in the D2-D6-brane configuration 
on the Eguchi-Hanson manifold 
in type-IIA string theory.
Branes are extended along directions denoted by $\circ$, and are not extended
along directions denoted by $-$. 
}
\label{tab:brane-ALE-wall}
\end{table}

Next, let us discuss 
a magnetic domain wall
in the Hanany-Witten configuration 
in type-IIB string theory
\cite{Eto:2004vy}.
 In this case, 
the position of the D3-brane 
at the $x^7$-coordinate depends on 
the $x^1$ coordinate 
for a magnetic domain wall.
Around the domain wall at $x^1=x_0$, 
they move from one D5-brane to the other 
D5-brane. Here, we consider the thin-wall limit for simplicity.
In Fig.~\ref{fig:easy-axis-wall} (c), they are represented by D3$^*$.
However, they can end on no D5 brane and must be bent to the $x^4$-direction to join to each other by creating a segment represented by D3$'$. 
In Fig.~\ref{fig:easy-axis-wall} (d), 
the same configuration is shown 
with plotting
the $x^1$ direction and 
suppressing the $x^4$ direction.
The brane configuration is 
summarized in Table \ref{tab:walls} 
.

\begin{table}[h]
\begin{center}
\begin{tabular}{c|cccccccccc}
\hline
&$x^0$&$x^1$&$x^2$&$x^3$&$x^4$&$x^5$&$x^6$&$x^7$&$x^8$&$x^9$\\
\hline
$\NC$ $\rm D3$&$\circ$&$\circ$&$\circ$&$\circ$&$-$&$-$&$-$&$-$&$-$&$-$ \\
\hline
D3$'$&$\circ$&$-$&$\circ$&$-$&$\circ$&$-$&$-$&$\circ$&$-$&$-$ \\
\hline
D3$^{\ast}$&$\circ$&$-$&$\circ$&$\circ$&$-$&$-$&$-$&$\circ$&$-$&$-$ \\
\hline
$\NF$ $\rm D5$&$\circ$&$\circ*$&$\circ*$&$-$&$\circ$&$\circ$&$\circ$&$-$&$-$&$-$ \\
\hline 
$2\; {\rm NS}5$&$\circ$&$\circ$&$\circ$&$-$&$-$&$-$
&$-$&$\circ$&$\circ$&$\circ$ \\
\hline
\end{tabular}
\end{center}
\caption{The domain wall in the Hanany-Witten brane configuration: Branes are extended along 
directions denoted by $\circ$, and are not extended
along directions denoted by $-$. 
\label{tab:walls}
}
\end{table}

\subsection{Chiral soliton lattice phase 
with easy-axis potential} \label{sec:CSL-EA}

\begin{figure}
\begin{center}
\begin{tabular}{cc}
\includegraphics[width=5cm,clip]{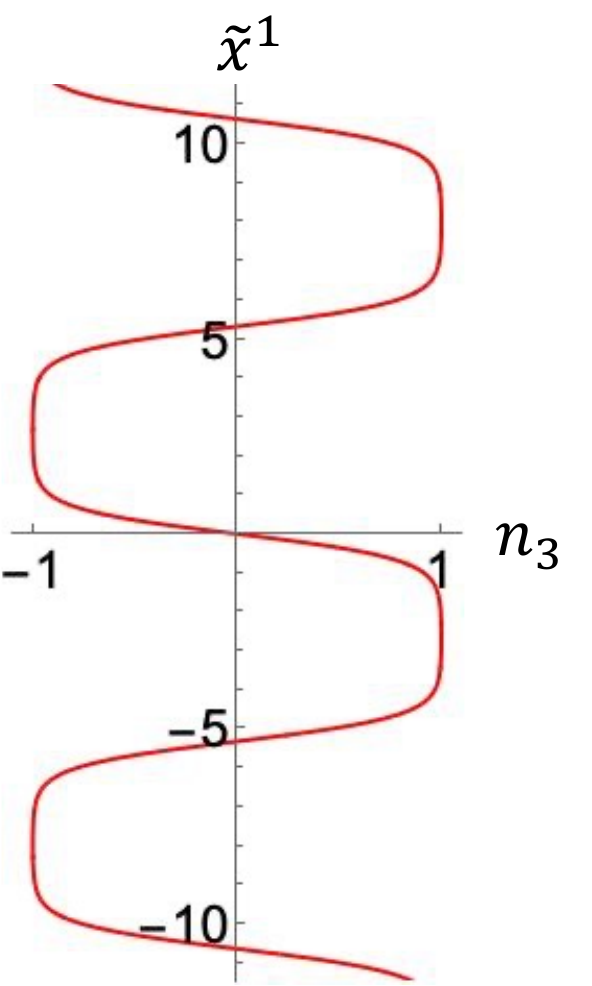} 
& \includegraphics[width=7cm,clip]{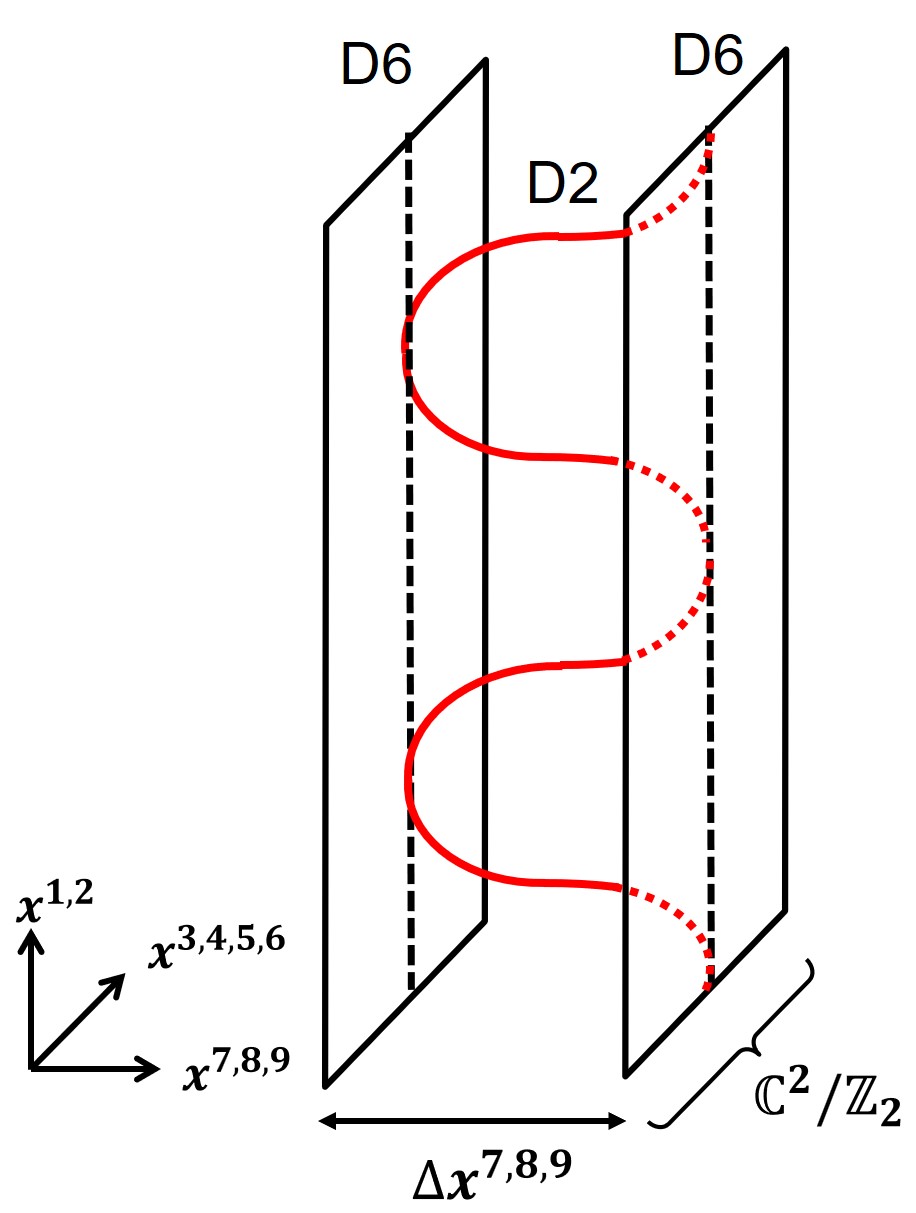}
\\
 a) &  b)  
\end{tabular}
\caption{
a) A CSL solution 
with the easy-plane potential 
as the ground state ($\kappa=1.0, m=1.3$).
b)  A snaky D2-brane 
corresponding to a). 
\label{fig:easy-axis-CSL}
} 
\end{center}
\end{figure}

We now discuss inhomogeneous ground states. 
Here, we give the condition 
that the CSL  
is the ground state 
instead of uniform configurations. 
As implied by the energy difference between the single soliton and the vacuum (uniform) configuration \eqref{eq:Energy-diff_kink-vac_easy-axis}, the kink energy can be lower than the vacuum energy.
In such a case, solitons 
are created in the vacuum,
and eventually they form a CSL, 
an array of kinks and anti-kinks.
Thus, a CSL is the ground state if
\begin{equation}
    4\left(2 m^{2}-\kappa^{2}\right)<\kappa^{2} \pi^{2} .
    \label{eq:easy-axis}
\end{equation}
The CSL solutions are given in terms of the Jacobi amplitude function as 
\begin{equation}
    f_\text{CSL}=\pm \mathrm{am}\left(\frac{\sqrt{2m^2-\kappa^2}}{\lambda}\tilde{x}^1 + X, \lambda \right) + \frac{\pi}{2} \ .
    \label{eq:CSL_easy-axis}
\end{equation}
It would be worth noting that Eq.~\eqref{eq:CSL_easy-axis} solves the Euler-Lagrange equation obtained from the energy functional for any $\lambda\in(0,1]$, but does not satisfy the BPS equation \eqref{eq:BPS-eq-kink}, except for the case $\lambda=1$ where it reduces to the single-kink solution \eqref{eq:kink_easy-axis}.\footnote{ 
Thus, the solution is 
non-BPS and breaks all SUSY.}
The solution \eqref{eq:CSL_easy-axis} with the plus (minus) sign is a monotonically increasing (decreasing) function of $\tilde{x}^1$. 
To obtain the ground state, the phase $\phi$ should be taken as Eq.~\eqref{eq:phase_GS_easy-axis}. In addition, the modulus $\lambda$ for the ground state is determined through
\begin{equation}
    2\mathrm{E}(\lambda) = \frac{|\kappa|\pi}{\sqrt{2m^2-\kappa^2}}\lambda
\end{equation}
with the elliptic integral of the second kind $\mathrm{E}(\lambda)$, which can be derived from $dE[f_\text{CSL}]/d\lambda=0$.
Fig.~\ref{fig:easy-axis-CSL} shows a CSL ground state 
for the easy-plane potential. The figure a) is a plot of the CSL solution, 
and b) is a schematic plot for a shape 
of the D2-brane in the D2-D6-ALE system, 
which may be called a snaky D-brane.

\subsection{Helimagnetic phase}\label{sec:helimagnetic}

When the relation $2m^2 = \kappa^2$ holds, 
the total potential term vanishes, 
as can be seen in Eq.~(\ref{eq:potential}), and 
the energy per unit length can be written as
\begin{equation}
    E[f]=\frac{1}{2}\int d\tilde{x}^1\[ \left\{\tilde{\partial}_1 f + \kappa \sin\left(\tilde{\vartheta} + \phi \right)\right\}^2 - \kappa^2\sin^2\left(\tilde{\vartheta} + \phi \right) \] + \text{const.}
\end{equation}
In this case, domain walls do not exist.
However, the ground state is uniformly modulated 
due to the DM interaction.
Since the BPS equation is given by
\begin{equation}
    \tilde{\partial}_1 f =- \kappa \sin\left(\tilde{\vartheta} +  \phi \right) \ ,
\end{equation}
the solution for the ground state is 
\begin{equation}
    \left\{
     \begin{array}{l}
    f = \pm \kappa  \tilde{x}^1
    \\
    \phi = - \tilde{\vartheta} \mp  \frac{\pi}{2} 
     \end{array}
    \right. \ .
\end{equation}
Thus, in this case, the phase varies linearly.

\subsection{Chiral soliton lattice phase with easy-plane potential}
\label{sec:CSL-EP}

Next, we consider the case of the easy-plane potential. 
A domain wall with the easy-plane potential is not topologically stable in the sense that both of the boundaries are connected and degenerated.
Such an unstable domain wall should decay into a vacuum state when the DM interaction is absent.
However, in the presence of the DM interaction, such domain walls can be energetically stable. Indeed, in our setting, the ground state is always given by a CSL composed of such unstable domain walls and anti-domain walls.

\begin{figure}
\begin{center}
\begin{tabular}{ccc}
\includegraphics[width=4.5cm,clip]{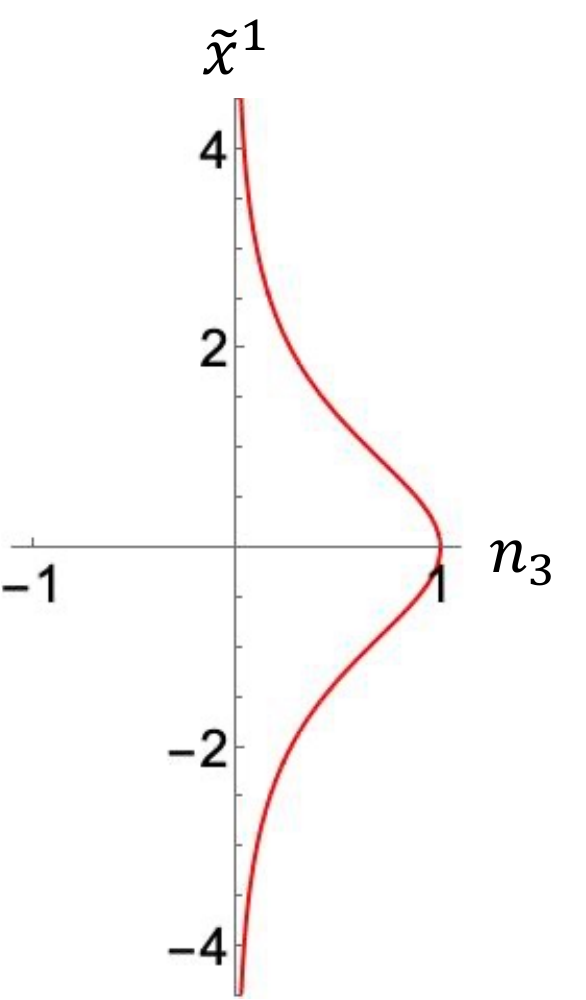} & 
\includegraphics[width=4.5cm,clip]{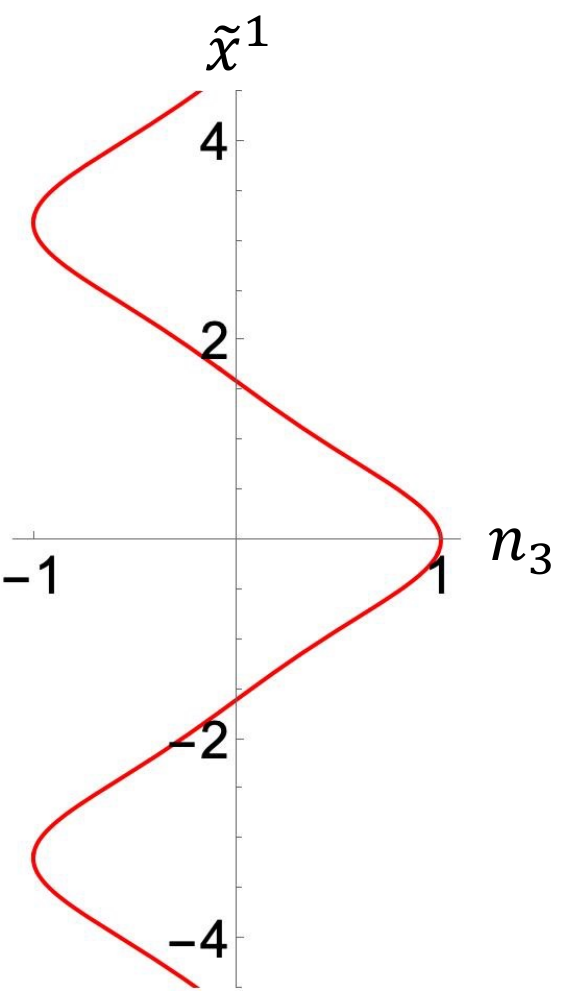} & 
\hspace{-0.8cm}
\includegraphics[width=7cm,clip]{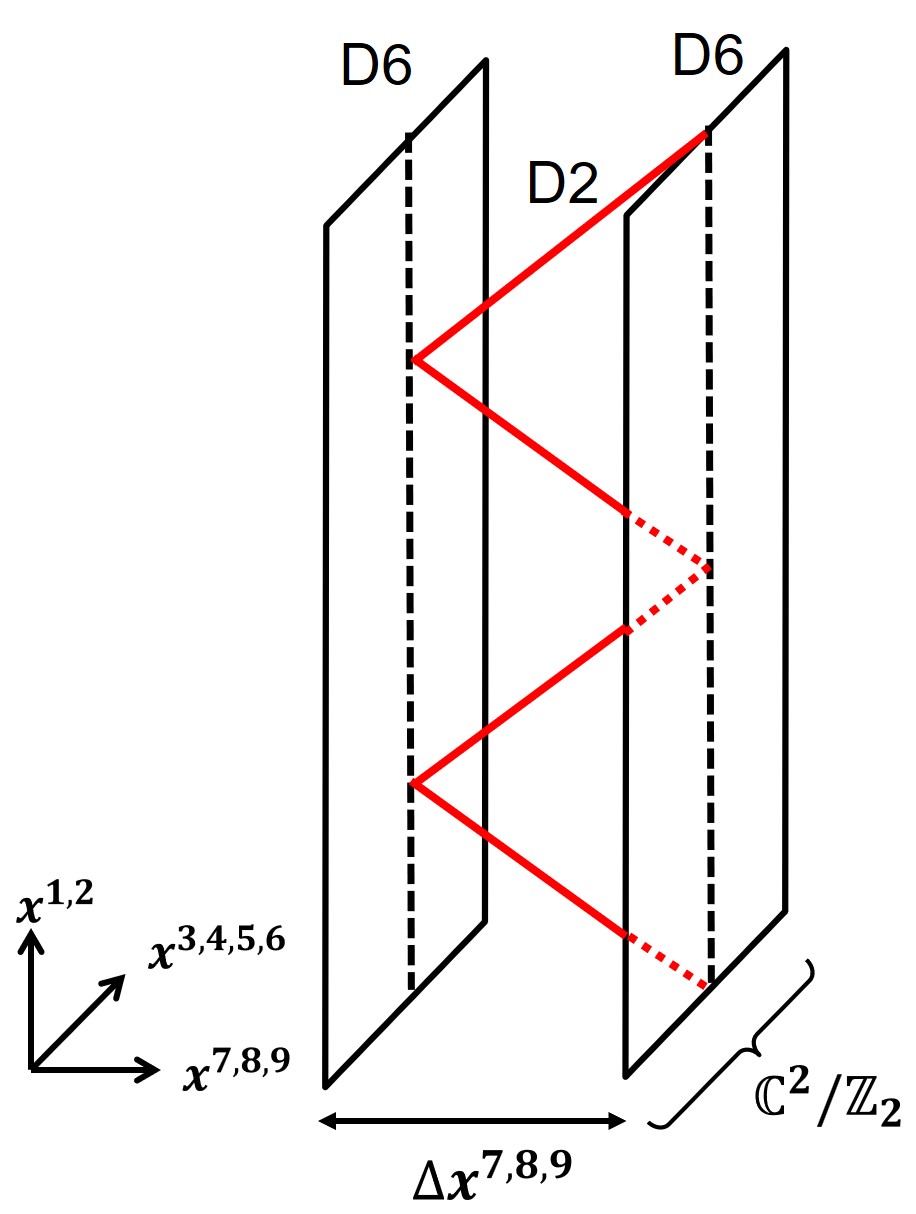} 
\\
 a) &  b)  & c)
\end{tabular}
\caption{a) A single  
domain wall and b) chiral soliton lattices with 
the easy-plane potential as the ground state ($\kappa = 1.2, m = 0.1$). 
c) A zigzag D2-brane corresponding to b).
\label{fig:easy-plane-CSL}}
\end{center}
\end{figure}

For the easy-plane case, the energy per unit length in the $\tilde{x}^2$-direction can be defined as
\begin{align}
    E[f]
    &=\int d\tilde{x}^1 \left[\frac{1}{2}\left(\tilde{\partial}_{1} f\right)^{2}+\kappa \sin (\tilde{\vartheta} +  \phi) \tilde{\partial}_{1} f +\left(\frac{\kappa^{2}}{2}-m^{2}\right) \cos ^{2} f \right] + E_\text{vac}^\text{e.p.} \ ,
    \label{eq:chiral-sG_energy_easy-plane}
\end{align}
where $E_\text{vac}^\text{e.p.}$ denotes the vacuum energy with the easy-plane potential.
Let us begin with studying a single kink solution.
The (anti-)BPS equation for a single kink is given by
\begin{equation}
    \tilde{\partial}_{1} f=\pm \sqrt{\kappa^{2}-2 m^{2}} \cos f
\end{equation}
This equation can be solved by
\begin{equation}
    f_{\text{kink}}=\pm 2 \arctan \left[\tanh \left(\frac{\sqrt{\kappa^{2}-2 m^{2}}}{2} \tilde{x}^{1} + X\right)\right]
    \label{eq:kink_easy-plane}
\end{equation}
where $X$ is a position moduli parameter. For the plus (minus) sign for an (anti-)BPS kink, the function \eqref{eq:kink_easy-plane} is a monotonically increasing (decreasing) function of $\tilde{x}^1$. 
Among the kink solutions, the lowest energy is attained when $\phi$ is taken as Eq.~\eqref{eq:phase_GS_easy-axis}, as the same with the easy-axis case.
Fig.~\ref{fig:easy-plane-CSL} a) shows a single kink solution in the easy-plane potential.

As the case of the easy-axis potential, the kink can have negative energy. 
In such a case, the kink does not decay to the vacuum.
The difference between the kink energy and vacuum energy is
\begin{equation}
    E\left[f_\text{kink}\right]-E_\text{vac}^\text{e.p.}=2 \sqrt{\kappa^{2}-2 m^{2}}-|\kappa| \pi.
\end{equation}
If the kink energy is less than 
that of uniform state (vacuum), i.e., 
\begin{equation}
    4\left(\kappa^{2}-2 m^{2}\right)<\kappa^{2} \pi^{2}
    \label{eq:easy-plane}
\end{equation}
then, a CSL, an array of kinks and anti-kinks, is the ground state. 
Since $\pi^{2}>4$, however, this inequality always holds. 
Thus, for the easy-plane potential, 
the ferromagnetic state (vacuum) is unstable, 
and the CSL phase is always the ground state for 
 any $m^{2} \in\left[0, \kappa^{2}/2\right)$. 

In the case of the easy-plane potential, the solution describing a CSL is given by
\begin{equation}
    f_\text{CSL}=\pm \mathrm{am}\left(\frac{\sqrt{\kappa^2 -2m^2}}{\lambda} \tilde{x}^1 + X, \lambda \right)
\end{equation}
with a position moduli parameter $X$.
Similar to the easy-axis case, the modulus $\lambda$ providing the ground state is determined through
\begin{equation}
    2\mathrm{E}(\lambda) = \frac{|\kappa|\pi}{\sqrt{\kappa^2-2m^2}}\lambda \ .
\end{equation}
Fig.~\ref{fig:easy-plane-CSL} b) shows 
a plot of the CSL solution with the easy-plane potential, 
which is rather zigzag 
compared with 
that with easy-axis potential.
This  can be identified with 
a D2-brane configuration for a CSL, 
as schematically drawn in 
Fig.~\ref{fig:easy-plane-CSL} c). 
The D2-brane touches the D6-branes at a shorter range than one in the easy-axis potential case shown 
in Fig.~\ref{fig:easy-axis-CSL}.

\section{Magnetic Skyrmions as D-branes}
\label{sec:skyrmions}

In this section, we discuss topologically excited states of two codimensions, that is, magnetic skyrmions and domain-wall skyrmions in chiral magnets. 
In our model, these are excited states on top of the easy-axis ferromagnetic ground state. 
In Subsec.~\ref{sec:skyrmion}, we numerically construct magnetic skyrmions and find that 
they are represented by D1-branes (fractional D0-branes) 
in the Hanany-Witten brane configuration (D2-D6-ALE system) 
in type-IIB(A) string theory.
In Subsec.~\ref{sec:DW-skyrmion}, 
we numerically construct explicit solutions of 
domain-wall skyrmions, and see the shape to the color D-brane 
in the visinity of a skyrmion.

\subsection{Magnetic skyrmions} \label{sec:skyrmion}

In this subsection, 
we show that magnetic skyrmions 
can be described 
by D1-branes in the Hanany-Witten brane 
configuration in type-IIB string theory, 
or by D0-branes in the D2-D4-ALE system 
in type-IIB string theory.
The latter are in fact fractional D0-branes, 
that is D2 branes two of whose spatial directions wrap around the $S^2$ cycle blowing up 
the orbifold singularity ${\mathbb C}^2/{\mathbb Z}_2$.
\begin{figure}[h]
\centering
\begin{tabular}{cc}
     \includegraphics[width=7cm,clip]{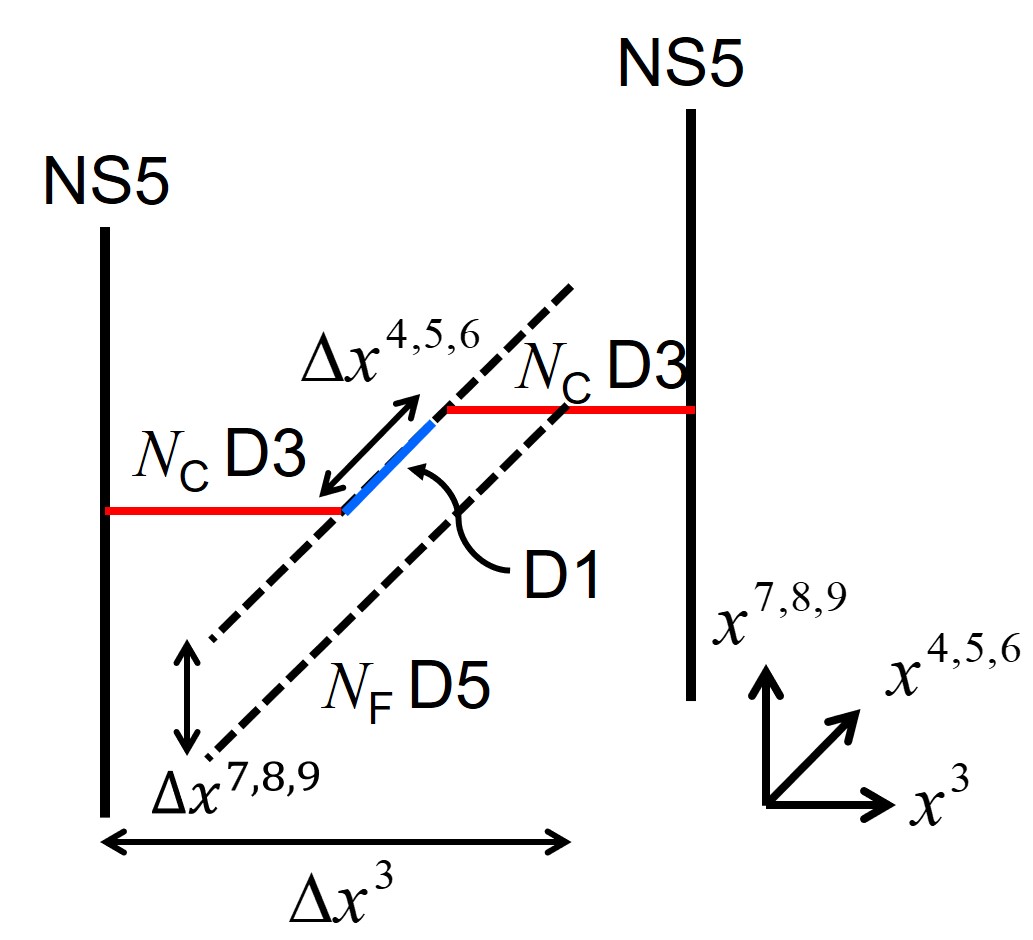} 
     &  
    \includegraphics[width=7cm,clip]{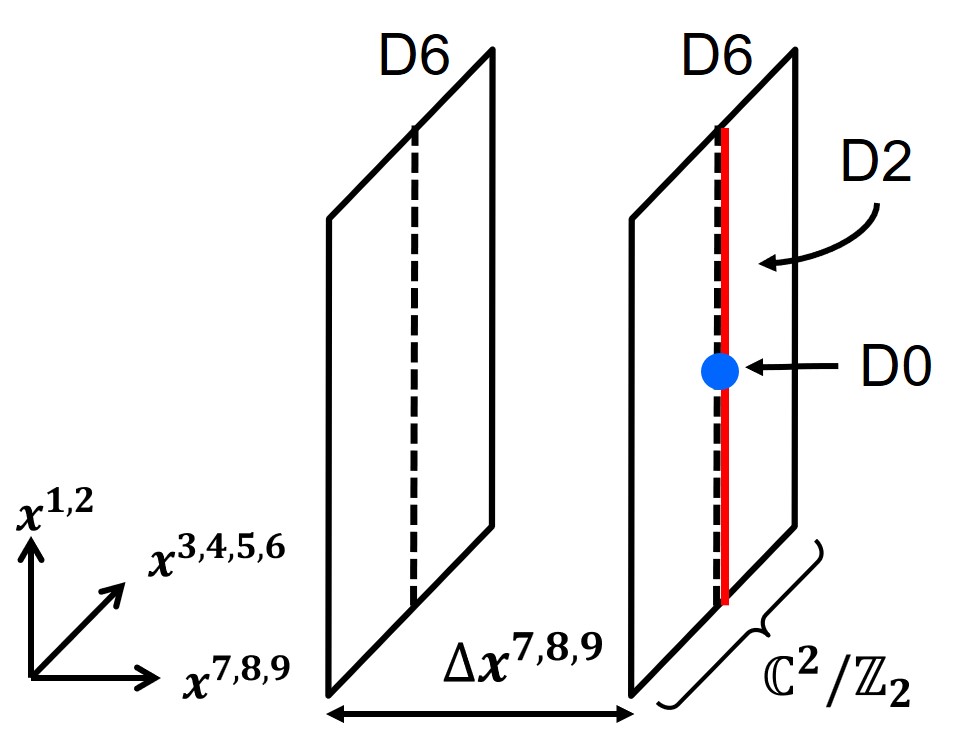} 
    \\
    a) & b) 
\end{tabular}
    \caption{D-brane configurations for a magnetic skyrmion 
    in a) the Hanany-Witten brane configuration and b) 
    the D2-D6-ALE system. 
    A magnetic skyrmion is realized 
    by a) a D1-brane and b) a fractional D0-brane, respectively.
    \label{fig:D-brane-skyrmion}}
\end{figure}
\begin{figure}[h]
\centering
     \includegraphics[width=1.0\columnwidth]{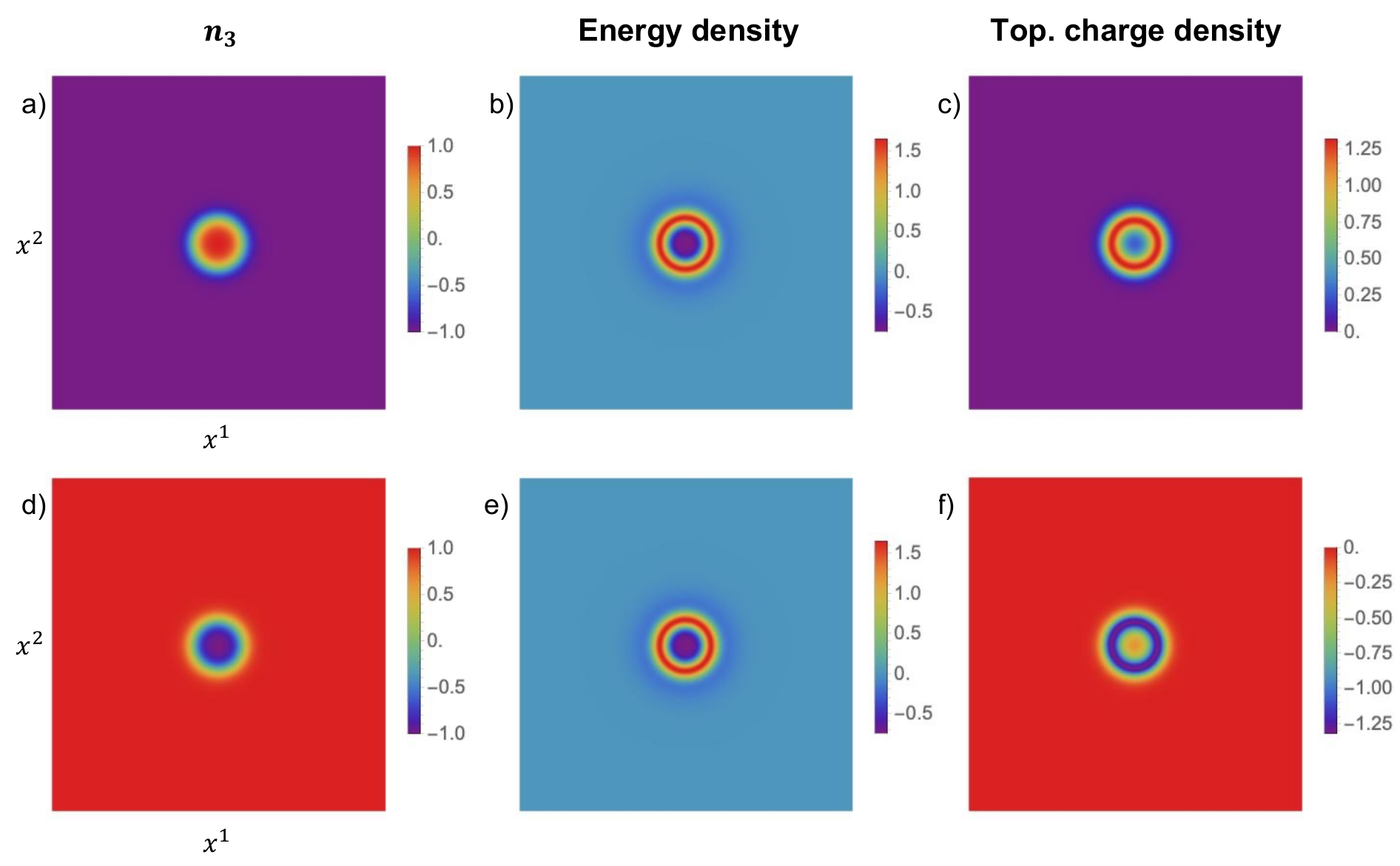} 
    \caption{Skyrmion and anti-skyrmion with the easy-axis potential ($\kappa=1.0, m^2=2.0, \vartheta=0$). 
    The top panels represent quantities of a skyrmion with the boundary condition $n_3=1$, and the bottom panels do of an anti-skyrmion with the boundary condition $n_3=-1$.
    The panels a) and d) show the value of $n_3$; b) and e) energy density; c) and f) topological charge density. 
    }
    \label{fig:skyrmion-solution}
\centering
     \includegraphics[width=1.0\columnwidth]{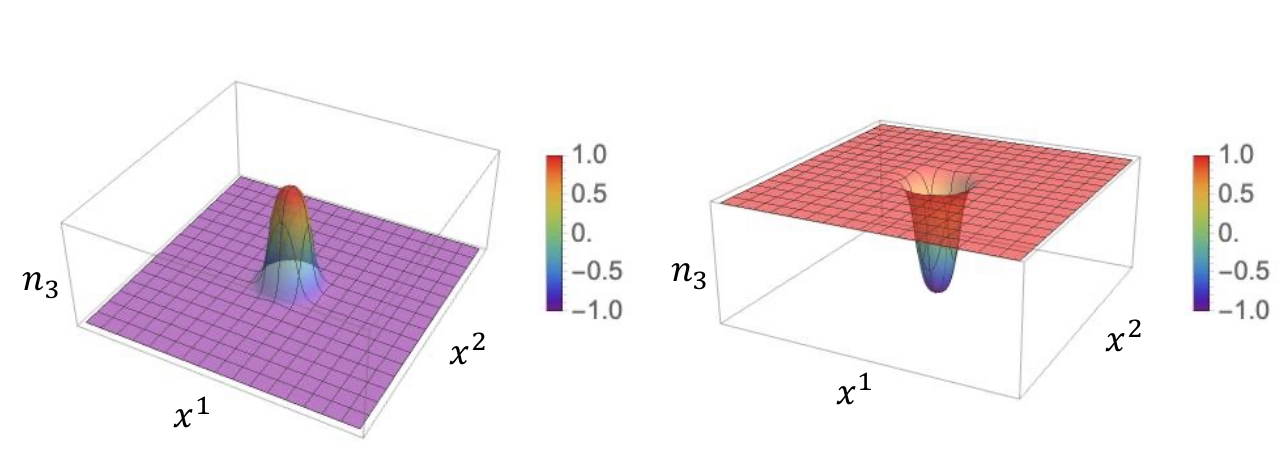} 
    \caption{Shape of a D2-brane for a single skyrmion (left)
 and anti-skyrmion (right) with the easy-axis potential ($\kappa=1.0, m^2=2.0, \vartheta=0$)}
    \label{fig:skyrmion-solution-D-brane}
\end{figure}

Let us start with the case
in the absence of the DM term
and masses of scalar fields 
($\Delta x^{7,8,9} = 0$).
In such a case, 
vortices were previously realized 
as 
D1-branes in the Hanany-Witten brane 
configuration, 
or as D0-branes in the D2-D6 system,
as in Fig.~\ref{fig:D-brane-skyrmion} a) or b), respectively  
\cite{Hanany:2003hp}.
These vortices are non-Abelian vortices in general \cite{Hanany:2003hp,Auzzi:2003fs,Eto:2005yh,Eto:2006cx,Shifman:2007ce,Shifman:2009zz,Eto:2004rz}. 
For $\NF = \NC \equiv N$ they are local non-Abelian vortices with non-Abelian orientational moduli 
${\mathbb C}P^{N-1}$, 
where the case of $N=1$ corresponds to 
Abrikosov-Nielsen-Olesen (ANO) 
vortices \cite{Abrikosov:1956sx,Nielsen:1973cs} 
in superconductors.
For $\NF >\NC$ such as our concern 
of $\NC=1, \NF=2$, 
they are semi-local vortices 
\cite{Vachaspati:1991dz,Achucarro:1999it,
Shifman:2006kd,Eto:2007yv}
having size and phase moduli 
in addition to non-Abelian orientational moduli.

In the strong gauge coupling limit, 
$\Delta x^3 \sim 1/g^2 = 0$, 
these semi-local vortices ($\NF >\NC$) become 
lumps in a nonlinear sigma model, 
which still possess size and phase moduli.
In the case of 
$\NC=1, \NF=2$, they are 
${\mathbb C}P^1$ lumps.

If we turn on masses of scalar fields 
($\Delta x^{7,8,9} \neq 0$), 
semilocal vortices 
at finite gauge coupling $g$ 
shrink, 
the size modulus becomes zero, 
and they eventually 
 become ANO votices.
However, 
in the strong gauge coupling limit, 
lumps are not stable in such a mass deformation; 
they shrink to zero size 
and configurations become singular 
(called small lump singularity).

Now we introduce the DM interaction 
by turning on the background gauge field 
(\ref{eq:bg-gauge}) 
on the D5(D6)-branes 
in the Hanany-Witten configuration 
(the D2-D4-ALE system).
Then, the D5(D6)-branes become 
so-called magnetized D-branes \cite{Bachas:1995ik,Berkooz:1996km,Blumenhagen:2000wh,Angelantonj:2000hi,
Cremades:2004wa,Kikuchi:2023awm,Abe:2021uxb}, 
and the $SU(2)$ gauge symmetry on the brane is spontaneously broken to $U(1)$. 
This induces the DM interaction
on the D3(D2)-brane worldvolume field theory, 
where the $SU(2)$ flavor symmetry is broken to $U(1)$.
The Hamiltonian density is 
given in Eq.~(\ref{eq:generalized-DM}), 
and 
the energy can be written as
\begin{align}
    E&\equiv\int d^2x {\cal H}
    \notag\\
    &=
        \int d^2x\left[\frac{1}{2}\partial_k\n \cdot \partial_k\n 
        + \kappa \left\{\cos\vartheta ~\n\cdot (\nabla\times\n) - \sin\vartheta (\n\cdot \nabla n_3 - n_3 \nabla\cdot \n) \right\} 
        \right. 
        \notag\\
        &\hspace{8cm}\left.
        + \left(m^2-\frac{\kappa^2}{2}\right)(1-n_3^2) \right]
    + \text{const.}
\end{align}
where the second term is the induced DM interaction, and the third is the easy-axis potential.
This induced DM interaction prevents lumps from shrinking, which are nothing but magnetic skyrmions. Indeed, the Derick's scaling argument \cite{Ross:2022vsa} requires that for stable configurations of codimension two, the energy contribution from the DM interaction energy $E_\text{DM}$ and the potential energy $E_\text{pot}$ satisfy the relation
\begin{equation}
    E_\text{DM} + 2 E_\text{pot} =0 \ , 
    \label{eq:Derrick}
\end{equation}
implying that non-trivial skyrmion solutions can exist only when the DM interaction energy is a finite negative value.
The topological charge of the skyrmion, $\pi_2(S^2)$, is defined by
\begin{equation}
    Q=\frac{1}{8\pi}\int d^2x ~\varepsilon^{jk} \n \cdot \left(\partial_j\n \times \partial_k \n \right) \ .
\end{equation}
Even in the presence of the DM interaction, only one of either skyrmion ($Q=+1$) or anti-skyrmion ($Q=-1$) is stable depending on the vacuum: When the vacuum is 
$n_3 = - 1$, a skyrmion is stable and an anti-skyrmion is unstable; On the other hand, when the vacuum is $n_3 = + 1$, an anti-skyrmion is stable and a skyrmion is unstable.\footnote{
This fact does not seem to be widely known in condensed matter community. Usually, one has the Zeeman energy 
$B n_3$. In such a case, only either skyrmion or anti-skyrmion 
is stable because of the unique vacuum.
} 
This can easily be seen when we consider a rotational symmetric ansatz of the form
\begin{equation}
    \n = \left( \cos(\nu \theta + \gamma)\sin f(r),
    \sin(\nu \theta + \gamma)\sin f(r), \cos f(r)\right)
    \label{eq:skyrmion_ansatz}
\end{equation}
with the polar coordinates $(r, \theta)$, where $\nu\in \mathbb{Z}$ denotes a winding number, $\gamma\in[0,2\pi)$ is a constant describing the internal orientation of the skyrmion, and $f(r)\in[0,\pi]$ is a monotonic function satisfying the boundary conditions $\{f(0)=0, f(\infty) =\pi\}$ or $\{f(0)=\pi, f(\infty) =0\}$.
Then, the DM term can be written as
\begin{equation}
    E_\text{DM} = -\kappa \int d^2x  ~ \sin[(1-\nu)\theta - \vartheta - \gamma]\left(\frac{\nu \sin(2f)}{2r} + \partial_r f\right) \ .
\end{equation}
It indicates that the energy contribution from the DM interaction vanishes except for $\nu = 1$. 
If $E_\text{DM}=0$, no non-trivial configuration can satisfy the relation \eqref{eq:Derrick}. Therefore, stable axisymmetric configurations must have the winding number $\nu=1$.
In addition, by substituting the ansatz into the topological charge, one obtains
\begin{equation}
    Q=-\frac{\nu}{2}\left[\cos f(r)\right]_{r=0}^{r=\infty} = -\frac{\nu}{2}\left[n_3 (r)\right]_{r=0}^{r=\infty} .
\end{equation}
Therefore, a stable configuration with $\nu=1$ possesses $Q=\pm 1$ when the vacuum is $n_3=\mp 1$.

The Euler-Lagrange equation of this model is given by
\begin{equation}
    \partial_b^2 n_a 
    +2\kappa\left[\sin\vartheta (\partial_a n_3 - \delta_{a3}\partial_bn_b) + \cos\vartheta ~\varepsilon_{abc}\partial_b n_c\right]
    +(2m^2-\kappa^2)\delta_{a3}n_3 + \Lambda n_a = 0
    \label{eq:eom_skyrmion}
\end{equation}
where $\Lambda$ is a Lagrange multiplier.
Note that, different from the kink and CSL cases, the DM interaction contributes not only to the energy but also to the equation of motion.
We numerically solve this equation using a conjugate gradient method with the fourth-order finite difference approximation. Our simulations are performed on a grid with $201\times 201$ lattice points and a lattice spacing $\Delta = 0.1$.
For the initial input, we employed the rotationally invariant configuration \eqref{eq:skyrmion_ansatz} with $\nu=1, \gamma = 0$, and a smooth monotonic function $f(r)$ satisfying the boundary condition.
In Fig.~\ref{fig:skyrmion-solution}, 
we give numerical solutions for 
a single magnetic (anti-)skyrmion 
with the easy-axis potential and the Bloch-type DM interaction, i.e., $\vartheta=0$.
As one can see from the energy density plots, 
the stable (anti-)skyrmion is in a shape 
of a domain wall ring,\footnote{A domain wall ring as a skrymion was found in the context of 
the baby skyrme model \cite{Kobayashi:2013ju}.
} 
inside which the energy density is negative 
due to the DM interaction.
Fig.~\ref{fig:skyrmion-solution-D-brane} shows that 
the color D2(D3)-brane worldvolume is bent and touches to the flavor D6(D5)-brane on the opposite side around the position of the D0(D1)-brane 
corresponding to the magnetic skyrmion 
in type IIA(B) string theory.

Since there is repulsion between skyrmions, multiple skyrmion states are always unstable, as the same with the magnetic skyrmions with a Zeeman interaction \cite{Shnir2018}.\footnote{
The presence of the Zeeman interaction $B n_3$ changes 
only the exponent of the asymptotic behavior, and thus 
only the strength of repulsion is 
different.
}

\subsection{Domain-wall Skyrmions}\label{sec:DW-skyrmion}

The ferromagnetic phase admits both magnetic domain walls and magnetic 
skyrmions. 
When they coexist, they feel attraction. 
Thus, a magnetic skyrmion should be absorbed into a domain wall.
The final state is a stable composite state called 
a domain-wall skyrmion
\cite{Nitta:2012xq,Kobayashi:2013ju,Ross:2022vsa}. 
In such a situation, 
the magnetic skyrmion is realized as a sine-Gordon soliton in the domain-wall effective field theory 
which is a sine-Gordon model 
with a potential term induced by the DM term \cite{Ross:2022vsa}.

\begin{figure}[h]
\centering
     \includegraphics[width=0.6\columnwidth]{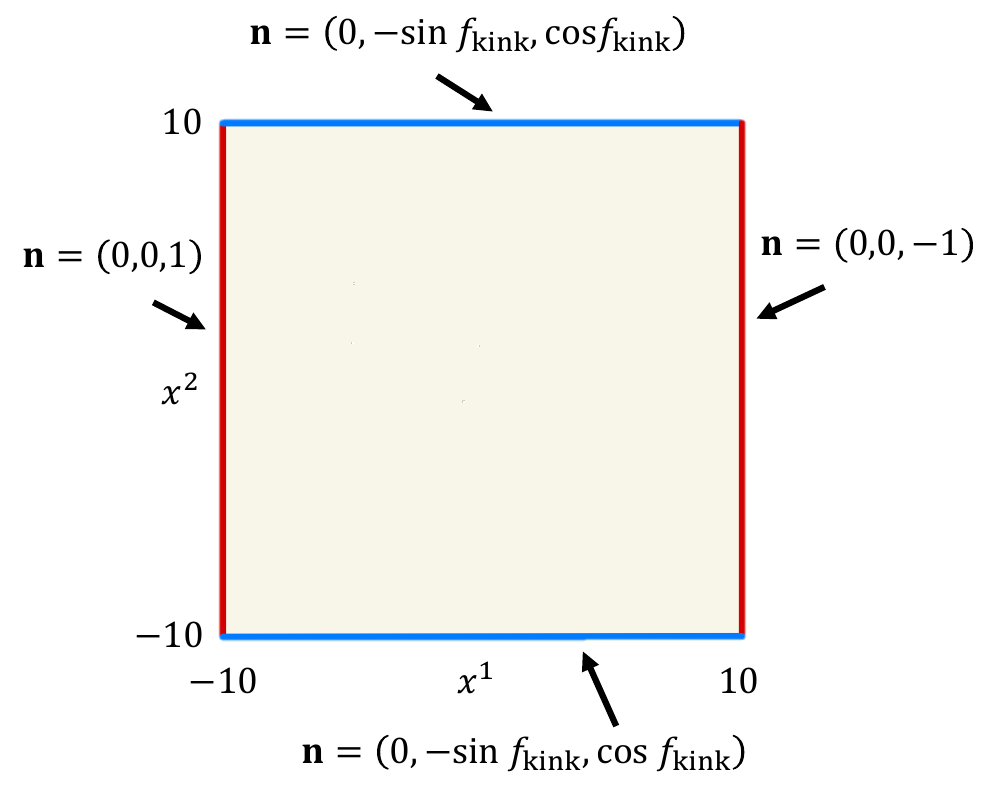} 
    \caption{Boundary condition used in the numerical simulation for the domain-wall skyrmions with the parameter $\kappa > 0, \vartheta=0 $.}
    \label{fig:DW-Skyrmion_BC}
\end{figure}
\begin{figure}[h]
\centering
     \includegraphics[width=1.0\columnwidth]{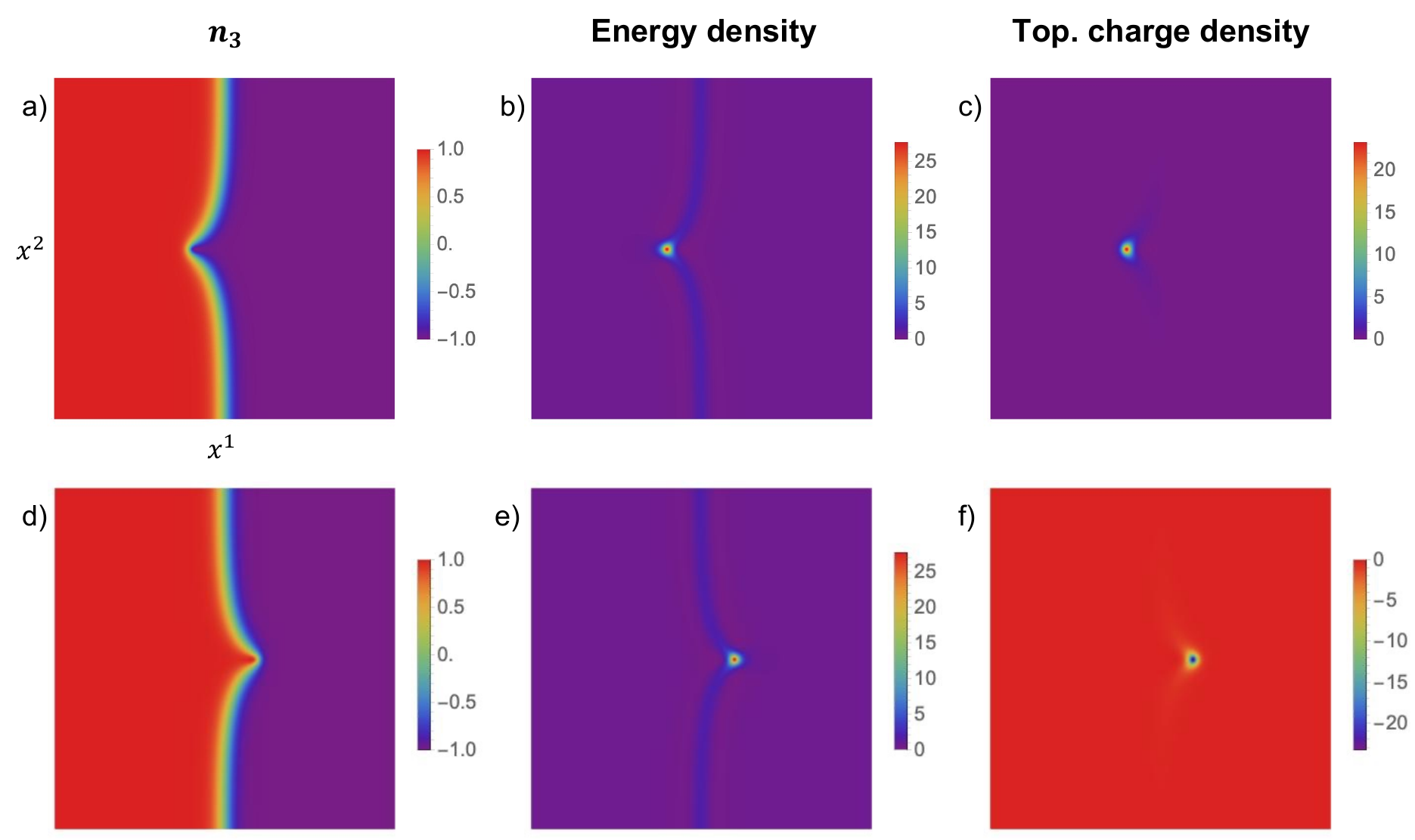} 
    \caption{Domain-wall skyrmion and 
    domain-wall anti-skyrmion with the easy-plane potential ($\kappa=1.0, m^2=2.0, \vartheta=0$).  The top panels represent quantities of a domain-wall skyrmion, and the bottom panels do of a domain-wall anti-skyrmion.
    The panels a) and d) show the value of $n_3$; b) and e) energy density; c) and f) topological charge density. a) $n_3$, b) energy density, c) topological charge density.}
    \label{fig:DW-skyrmion-solution}
\end{figure}
\begin{figure}[thb]
\begin{center}
\begin{tabular}{cc}
\includegraphics[width=8cm,clip]{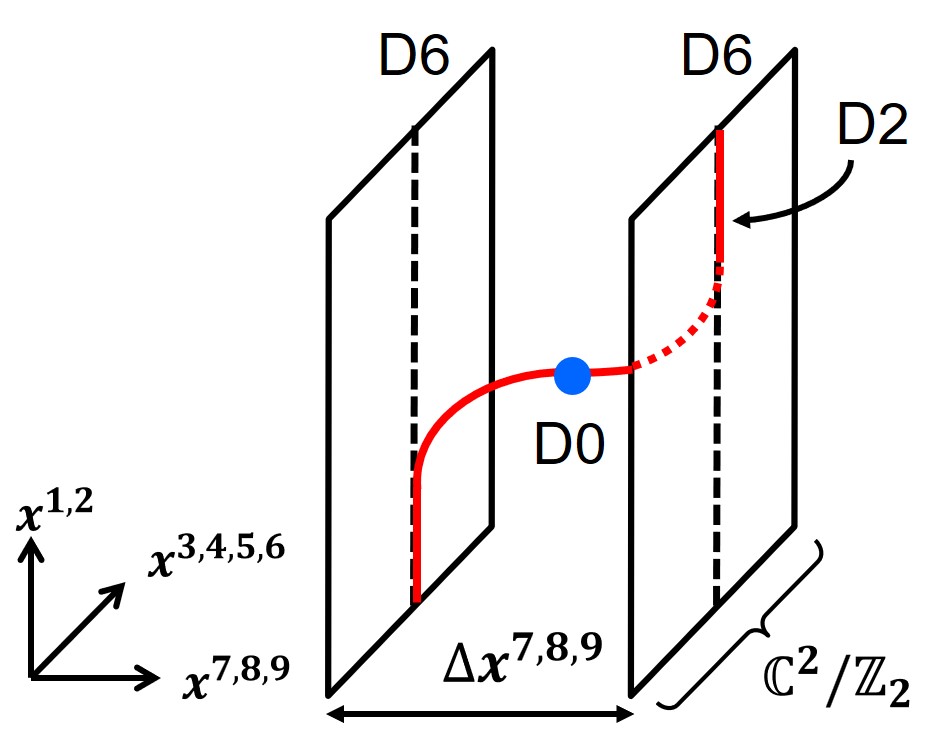} 
& \includegraphics[width=8cm,clip]{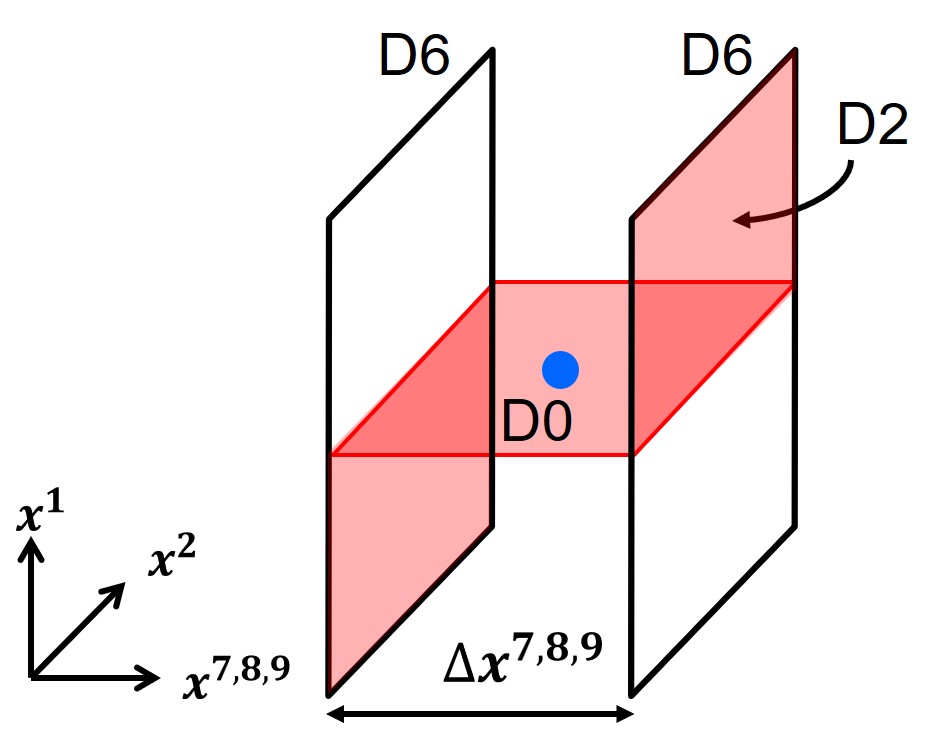} \\
    (a) & (b) 
\end{tabular}
\end{center}
\caption{
A kinky D2-D0-brane bound state 
in the D2-D6-ALE system in type-IIA string theory describing 
a domain-wall skyrmion. 
A similar configuration can be considered in the Hanany-Witten brane configuration in type-IIB string theory.
\label{fig:kinky-brane-DWskyrmion}
}
\centering
     \includegraphics[width=1.0\columnwidth]{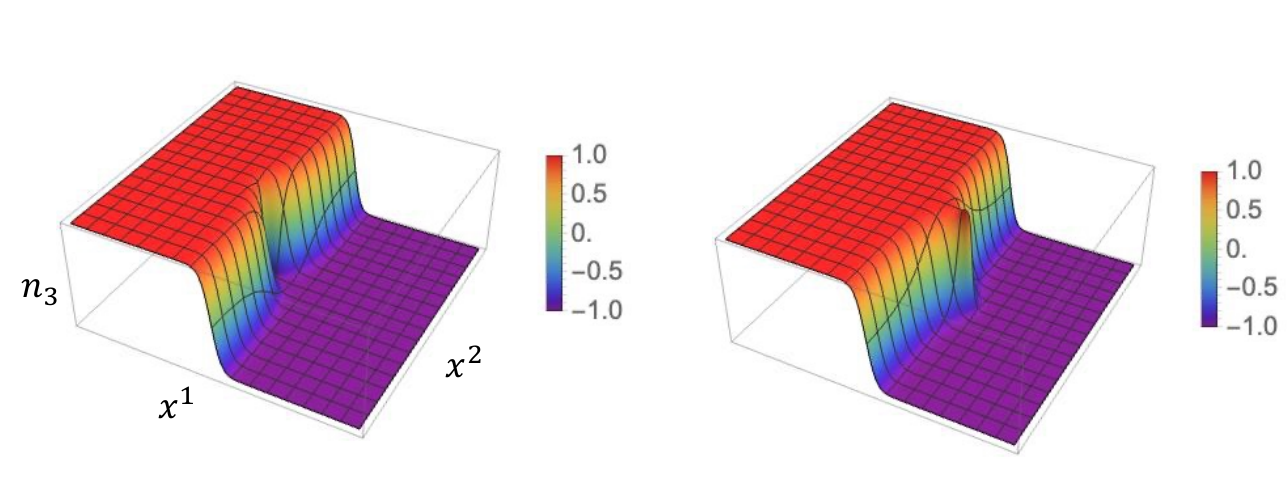} 
    \caption{Shapes of the D2-brane for a domain-wall skyrmion (left) and anti-skyrmion (right) with the easy-axis potential ($\kappa=1.0, m^2=2.0, \vartheta=0$). 
    The kinky shape of the D2-branes is pulled or pushed 
    due to the D0-brane.
    }
    \label{fig:DW-skyrmion-solution-D-brane}
\end{figure}

To obtain explicit solutions of domain-wall skyrmions, we numerically solve the equation of motion in Eq.~\eqref{eq:eom_skyrmion} with the same method used in the last subsection. We run our simulations by replacing the spatial domain $[-10,10]\times[-10,10]\subset \mathbb{R}^2$ with a grid with $201\times 201$ lattice points and lattice spacing $\Delta=0.1$.
We impose the Dirichlet boundary conditions that we assign a different vacuum to each boundary in the $x^2$-direction and the lowest energy single-kink solution with the easy-axis potential in Eq.~\eqref{eq:kink_easy-axis} to the boundaries in the $x^1$-direction.  
In Fig.~\ref{fig:DW-Skyrmion_BC}, 
we show a concrete example of the boundary condition.
In the bulk, we employ the following configuration as an initial input by choosing parameters compatible with the boundary conditions:
\begin{equation}
    \n = \left( \cos\phi(x^2)\sin f_\text{kink}(x^1), 
    \sin\phi(x^2)\sin f_\text{kink}(x^1),
    \cos f_\text{kink}(x^1) \right)
\end{equation}
where $f_\text{kink}$ is the single kink solution for the easy -axis case in Eq.~\eqref{eq:kink_easy-axis} with the moduli parameter $X=0$, and 
\begin{equation}
    \phi = 4\arctan e^{c x^2} - \vartheta \mp \frac{\pi}{2}
\end{equation}
with a real parameter $c$. 
In Fig.~\ref{fig:DW-skyrmion-solution}, we present numerical solutions 
for domain-wall (anti-)skyrmions. The upper figures show a single skyrmion 
on the wall 
and the lower figures show a single anti-skyrmion on the wall.
It is important to emphasize that both skyrmion and anti-skyrmion are stable.
We can observe that the domain wall  
is bent in the vicinity of the (anti-)skyrmion, which is consistent with Ref.~\cite{KBRBSK}.
The direction of the bending depends on the sign 
of the topological charge of the skyrmion.


In Fig.~\ref{fig:kinky-brane-DWskyrmion}, 
we schematically draw brane configurations for a domain-wall skyrmion, 
as a combination of those for a domain wall and magnetic skyrmion.
As a consequence of the bending of the domain wall in Fig.~\ref{fig:DW-skyrmion-solution},
 the D2-brane worldvolume forming a kink is pulled to either direction of the wall 
in the vicinity of a D0-brane corresponding 
to the magnetic skyrmion, 
as shown in Fig.~\ref{fig:DW-skyrmion-solution-D-brane}.

\section{Summary and Discussion}
\label{sec:summary}

We have given string theory construction for 
chiral magnets in terms of 
the Hanany-Witten brane configuration 
(consisting of D3, D5 and NS5-branes) in type-IIB string theory,
and the fractional D2 and D6 branes on the Eguchi-Hanson manifold 
in type-IIA string theory.
In both cases,   
the flavor branes are magnetized by 
a constant magnetic flux.
The $O(3)$ sigma model with 
the DM interaction describing chiral magnets 
is realized on 
the worldvolume of 
the color D-branes.
As summarized in Fig.~\ref{fig:phase-diagram}, 
we have found that the ground states 
are not uniform in general:
The ground state is either a ferromagnetic (uniform) state, 
a CSL phase with the easy-axis potential 
or the easy-plane potential, 
or the helimagnetic state.
A magnetic domain wall 
in the ferromagnetic phase is 
realized by a kinky D-brane.
In the CSL phase with the easy-axis (plane) potential, the uniform state is unstable 
because a single (non)topological 
domain wall 
has negative energy due to the DM interaction. 
Consequently, 
the color D-brane is 
snaky 
(zigzag) between 
the two separated flavor D-branes 
as in Fig.~\ref{fig:easy-axis-CSL} 
(\ref{fig:easy-plane-CSL}).
We also have constructed magnetic skyrmions 
realized as D1-branes (fractional D0-branes)
in 
the former (latter) configuration.
We have shown that the worldvolume of the host D2-brane is bent at 
the position of 
the D0-brane as the magnetic skyrmion 
and is touched to the other flavor D-brane,  
see Fig.~\ref{fig:skyrmion-solution-D-brane}.
Finally, we have constructed domain-wall skyrmions   
in the ferromagnetic phase. 
The domain-wall worldvolume 
is no longer flat in the vicinity of the (anti-)skyrmion and is pulled 
into a direction determined 
from the skyrmion topological charge 
as in Fig.~\ref{fig:DW-skyrmion-solution}.
Consequently, the D2-brane worldvolume  
is pulled  
in the vicinity of the D0-brane 
as in Fig.~\ref{fig:DW-skyrmion-solution-D-brane}.

Before closing this paper, 
let us discuss future directions.
In this paper, we have studied domain-wall skyrmions in the ferromagnetic phase. Recently, they are generalized to domain-wall bimerons and domain-wall skyrmion-chain in the CSL phase \cite{Amari:2023bmx}. D-brane configurations of these cases are natural extensions to be explored.

One of the most important directions 
may be the introduction of the Zeeman term $n_3$
induced by an applied magnetic field.
In such a case, 
a skyrmion lattice phase is also a possible ground state
in the phase diagram~\cite{doi:10.1126/science.1166767,doi:10.1038/nature09124,doi:10.1038/nphys2045},
where skyrmions have negative energy 
due to the DM term \cite{Rossler:2006,Han:2010by,Lin:2014ada,Ross:2020hsw}. 
Whether such a term can be introduced in brane configurations 
will be an interesting and important open question.

The three inhomogeneous  
ground states, the CSL phases with easy-axis (plane) potential
and the helimagnetic phase, are continuously connected or crossover.
On the other hand, 
the defect-type phase transition exists between 
the ferromagnetic phase to 
the easy-axis CSL phase \cite{deGennes1975,PhysRevB.101.214424}. 
Soliton formations of this transition 
were studied in Refs.~\cite{Eto:2022lhu,Higaki:2022gnw} 
in the framework of the chiral sine-Gordon model.
These studies should be extended to 
the case of chiral magnets, 
the $O(3)$ model with the DM term.

Domain walls and skyrmions 
are related by a Scherck-Schwarz dimensional reduction 
\cite{Eto:2004rz} or a T-duality 
\cite{Eto:2006mz,Eto:2007aw}
in string theory language, 
at least in the absence of the DM term.
It is not clear 
if this duality holds 
with the DM term. 
Physically, this is related 
to the aforementioned 
phase transition.

\begin{figure}[h]
\centering
     \includegraphics[width=0.6\columnwidth]{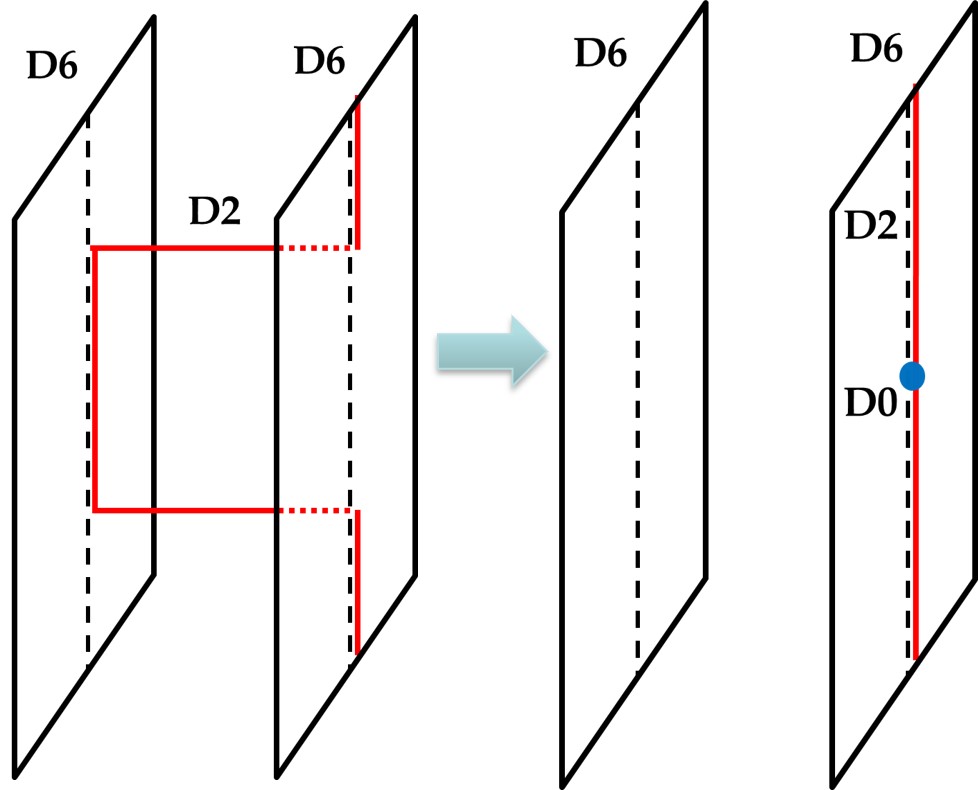} 
    \caption{A pair annihilation of kinky D2-brane and anti-kinky D2-brane resulting in the creation of D0-branes.}
    \label{fig:brane-annihilation}
\end{figure}

It is known that 
when a D$p$-brane and anti D$p$-brane pair annihilate, 
D$(p-2)$-branes are created 
as a consequence of a tachyon condensation 
\cite{Sen:1998sm,Sen:2004nf}. 
A kinky D-brane and anti-kinky D-brane can annihilate 
for instance at 
a phase transition from 
the CSL phase to ferromagnetic phase.
Locally this annihilation can be regarded as 
a D2-brane anti D2-brane pair annhilation as in 
Fig.~\ref{fig:brane-annihilation} (left). 
Consequently, there appear D0-branes afeter the pair annihilation 
as in Fig.~\ref{fig:brane-annihilation} (right). 
These are nothing but magnetic skyrmions.
A pair annihilation of domain wall and anti-domain wall resulting 
in the creation of magnetic skyrmions. 
This process was studied 
in the $O(3)$ model without the DM term
\cite{Nitta:2012kj,Nitta:2012kk} 
and Bose-Einstein condensates 
\cite{Takeuchi:2012ee,Takeuchi:2012ap,Nitta:2012hy,Takeuchi:2011uyb}. 
In these cases, the $U(1)$ moduli exist on the domain walls 
(see footnote \ref{footnote:U(1)}), and 
the creation rate of skyrmions depends on the relative phase.
The creation rate is maximized 
when the relative phase moduli is $\pi$, 
which is precisely the 
case with the DM term.

Let us make comments on the stability of configurations studied in this paper. 
In the nonlinear sigma model limit 
$g\to \infty$, the configurations are known to be stable.
At finite coupling $g$, the theory is a gauge theory. 
The stability analysis in this case 
has not been studied before and remains an important open question. 
Finally, in string theory, 
we expect the stability should be the same with that of the gauge theory, 
as far as we are considering the field theory limit. 
However, the most subtle point is that the background gauge 
field on the flavor D-branes breaks SUSY.  
As far as we turn on small $\kappa$, the configurations 
are expected to be stable but the full analysis of this problem is beyond the scope of this paper. 
The previous studies 
\cite{Bachas:1995ik,Berkooz:1996km,Blumenhagen:2000wh,Angelantonj:2000hi,
Cremades:2004wa,Kikuchi:2023awm,Abe:2021uxb}
on magnetized D-branes in simpler setups 
should be useful for this purpose.
We will come back to this problem in future. 

Generalizations of chiral magnets to the 
${\mathbb C}P^2$ model or more generally to the ${\mathbb C}P^{N-1}$ model were studied before 
\cite{Akagi:2021dpk,Akagi:2021lva,Amari:2022boe}
in which magnetic skyrmions were mainly investigated.
On the other hand, multiple domain walls 
were studied without the DM interaction in 
the ${\mathbb C}P^{N-1}$ model 
\cite{Gauntlett:2000ib,Tong:2002hi} and  
Grassmann model 
\cite{Isozumi:2004jc,Isozumi:2004va,Isozumi:2004vg}, for which 
multiple kinky D-brane configurations  were explored in Ref.~\cite{Eto:2004vy}.
Domain-wall skyrmions 
were also 
constructed without the DM interaction 
in parallel multiple walls in the ${\mathbb C}P^{N-1}$ model \cite{Fujimori:2016tmw} and 
a single non-Abelian domain wall in the Grassmann model 
\cite{Nitta:2015mma,Nitta:2015mxa}.
Furthermore,
domain wall junctions or networks
\cite{Eto:2005cp,Eto:2005fm,
Eto:2006bb,Eto:2007uc,
Eto:2020vjm,Eto:2020cys} 
and their D-brane configurations
\cite{Eto:2005mx} were also studied.
Introducing the DM interaction 
in these models will be interesting 
to explore because these models, at least the ${\mathbb C}P^2$ model,  can be experimentally realizable 
in laboratory experiments of ultracold atomic gases.

Finally, 
modulated ground states (vacua) 
were discussed in string theory 
\cite{Nakamura:2009tf,
Andrade:2015iyf,
Andrade:2017leb,
Andrade:2017cnc,
Cai:2017qdz} 
and relativistic 
field theory 
\cite{
Nitta:2017mgk,
Nitta:2017yuf,
Gudnason:2018bqb,
BjarkeGudnason:2018aij,
Musso:2018wbv}. 
These studies may be useful 
for analyzing various aspects of 
the modulated phases found in this paper,  
for instance phonons in the CSL.

\begin{acknowledgments}
We thank Ryo Yokokura and Tetsutaro Higaki for their useful comments.
This work is supported in part by 
 JSPS KAKENHI [Grants No. JP23KJ1881 (YA) and No. JP22H01221 (MN)], the WPI program ``Sustainability with Knotted Chiral Meta Matter (SKCM$^2$)'' at Hiroshima University.
 The numerical computations in this paper were run on the ``GOVORUN" cluster supported by the LIT, JINR.
 
\end{acknowledgments}



\begin{appendix}
	\section{The Dzyaloshinskii-Moriya interaction as a background gauge field}
 \label{sec:DM-from-BGG}

In this Appendix, we show that the $SU(2)$ gauged $O(3)$ nonlinear sigma model in Eq.~\eqref{eq:gauged_sigma_model} can be derived from the strong coupling limit of 
the $U(1) \times SU(2)$ gauged linear sigma model given in Eq.~\eqref{eq:U2_gauge_theory}, i.e.,
\begin{align}
\mathcal{L}=2\left({\cal D}_{\mu} \Phi\right)^{\dagger}\left({\cal D}^{\mu} \Phi\right) - \Phi^\dagger(\Sigma\mathbf{1}_2-M)^2\Phi
\label{eq:Lagrangian_app}
\end{align}
with
${\cal D}_{\mu} \Phi = \left(\pd_\mu -i a_{\mu} -\frac{i}{2} {A}_\mu \right)\Phi$ and 
a $U(1)$ auxiliary gauge field $a_\mu$.
Since the Lagrangian does not contain the kinetic term of $a_\mu$ and $\Sigma$ in the strong coupling limit, they are auxiliary fields. Similar to the case in which $A_\mu$ is absent given in Eq.~\eqref{eq:elimination_auxiliary}, one can eliminate the fields using the equations of motion as
\begin{equation}
    a_\mu = \frac{i}{2}\left\{(\mathscr{D}_\mu \Phi)^\dagger\Phi - \Phi^\dagger(\mathscr{D}_\mu \Phi) \right\},
    \qquad
    \Sigma= \Phi^\dagger M \Phi =mn_3
\end{equation}
with $\mathscr{D}_\mu \equiv \partial_\mu -\frac{i}{2} {A}_\mu$.
Substituting them into the Lagrangian in Eq.~\eqref{eq:Lagrangian_app}, we obtain
\begin{align}
 \mathcal{L} & =2\left\{ \left(\mathscr{D}_{\mu} \Phi\right)^\dagger\left(\mathscr{D}^{\mu} \Phi\right)+\left(\Phi^\dagger \mathscr{D}_{\mu} \Phi\right)^{2}\right\} - m^2(1-n_3^2) \ .
     \label{eq:Lag_Phi}
\end{align}
By expanding the expression, we can rewrite the Lagrangian as
\begin{equation}
    \begin{split}
        {\cal L} = &2 \left\{\partial_\mu\Phi^\dagger\partial^\mu \Phi +(\Phi^\dagger\partial_\mu \Phi) \right\}
        \\
        &
        +i\Phi^\dagger A_\mu\partial^\mu\Phi -i\partial^\mu\Phi^\dagger A_\mu\Phi -2i\Phi^\dagger A_\mu \Phi \Phi^\dagger\partial^\mu \Phi
       \\&+\frac{1}{2}\left\{(A_\mu^a)^2 - (\Phi^\dagger A_\mu \Phi)^2 \right\}
       -m^2(1-n_3^2) \ .
    \end{split}
        \label{eq:Lag_Phi_expand}
\end{equation}
Here, we have used a relation
\begin{equation}
    \Phi^\dagger A_\mu A^\mu \Phi 
    =A_\mu^a \left(A^\mu\right)^b \Phi \sigma_a\sigma_b \Phi 
    =A_\mu^a \left(A^\mu\right)^b ~\Phi (\delta_{ab}\mathbf{1}_2 +i \varepsilon^{abc}\sigma_c) \Phi = (A_\mu^a)^2 \ .
\end{equation}

We now compare this Lagrangian and the gauged nonlinear sigma model.
First, we can write the Dirichlet term as
\begin{align}
    \pd_\mu\n \cdot \pd^\mu \n 
    =& (\sigma_{a})_{\alpha\beta}(\sigma_b)_{\gamma\delta} ~\pd_\mu(\Phi^*_\alpha\Phi_\beta) \pd_\mu(\Phi^*_\gamma\Phi_\delta) 
    \notag\\
    =&(2\delta_{\alpha\delta}\delta_{\beta\gamma}-\delta_{\alpha\beta}\delta_{\gamma\delta})\pd_\mu(\Phi^*_\alpha\Phi_\beta) \pd_\mu(\Phi^*_\gamma\Phi_\delta) 
    \notag\\
    =&2\pd_\mu(\Phi^*_\alpha\Phi_\beta) \pd_\mu(\Phi^*_\beta\Phi_\alpha) 
    \notag\\
    =&4\left\{\partial_\mu\Phi^\dagger\partial^\mu \Phi +(\Phi^\dagger\partial_\mu \Phi) \right\} \ .
    \label{eq:Dirichlet_term}
\end{align}
The DM interaction can be rewritten as
\begin{align}
\bm{A}_\mu \cdot\left(\n \times \partial^\mu \n\right) 
& =\varepsilon^{a b c} A_\mu^{a} n_{b} \partial^\mu n_{c} 
\notag\\
& =\varepsilon^{a b c} A_\mu^{a} ~\Phi^\dagger \sigma_{b} \Phi ~\partial^\mu\left(\Phi^{\dagger} \sigma_{c} \Phi\right) 
\notag\\
& =i A_\mu^{a}~\Phi^\dagger\left(\sigma_{a} \sigma_{b}-\delta_{a b}\mathbf{1}_2\right) \Phi~\partial^\mu\left(\Phi^{\dagger} \sigma_{b} \Phi\right)   
\notag\\
& =i A^{a}_\mu\Phi^\dagger ~\sigma_{a} \sigma_{b} \Phi ~\partial^\mu\left(\Phi^{\dagger} \sigma_{b} \Phi\right) -i ~\partial^\mu\left(\Phi^{\dagger} {A}_\mu \Phi\right) 
\end{align}
where we have used $\varepsilon^{a b c} \sigma_{c}=-i\left(\sigma_{a} \sigma_{b}-\delta_{a b}\mathbf{1}_2\right)$. Using the relation $(\sigma_a)_{\alpha \beta}(\sigma_a)_{\gamma\delta} =(2 \delta_{\alpha \delta} \delta_{\beta \gamma}-\delta_{\alpha \beta} \delta_{\gamma \delta})$, one obtains
\begin{align}
\Phi^\dagger \sigma_{a} \sigma_{b} \Phi \Phi^\dagger \sigma_{b} \partial_\mu \Phi 
& =\left(\Phi^\dagger \sigma_{a}\right)_{\alpha} (\sigma_{b})_{\alpha \beta} \Phi_{\beta} ~\Phi_{\gamma}^\dagger (\sigma_{b})_{\gamma \delta} \partial_\mu \Phi_{\delta} 
\notag\\
& =\left(2 \delta_{\alpha \delta} \delta_{\beta \gamma}-\delta_{\alpha \beta} \delta_{\gamma \delta}\right)\left(\Phi^\dagger \sigma_{a}\right)_{\alpha} \Phi_{\beta}~ \Phi_{\gamma}^\dagger \partial_\mu \Phi_{\delta} 
\notag\\
& =2 \Phi^\dagger \sigma_{a} \partial_{k} \Phi-\Phi^\dagger \sigma_{a} \Phi ~\Phi^\dagger \partial_\mu \Phi 
\end{align}
and
\begin{align}
\Phi^\dagger \sigma_{a} \sigma_{b} \Phi ~\partial_\mu \Phi^\dagger \sigma_{b} \Phi
& =(\sigma_{b})_{\alpha \beta} (\sigma_b)_{\gamma \delta}\left(\Phi^\dagger \sigma_{a}\right)_{\alpha} \Phi_{\beta} \partial_\mu \Phi_{\gamma}^{\dagger} \Phi_{\delta} 
\notag\\
& =\left(2 \delta_{\alpha \delta} \delta_{\beta \gamma}-\delta_{\alpha \beta} \delta_{\gamma \delta}\right)\left(\Phi^\dagger \sigma_{a}\right)_{\alpha} \Phi_{\beta} \partial_\mu \Phi^\dagger \Phi_{\delta} 
\notag\\
& =2 \Phi^\dagger \sigma_{a} \Phi \partial_\mu \Phi^\dagger \Phi-\Phi^\dagger \sigma_{a} \Phi \partial_\mu \Phi^\dagger \Phi 
\notag\\
& =\Phi^{\dagger} \sigma_{a} \Phi \partial_\mu \Phi^\dagger \Phi \ .
\end{align}
Combining them, we can rewrite the DM interaction as
\begin{align}
\bm{A}_\mu \cdot\left(\n \times \partial^\mu \n\right) 
= & i\left\{2 \Phi^\dagger {A}_\mu \partial^\mu \Phi-\Phi^\dagger {A}_\mu \Phi \Phi^\dagger \partial^\mu \Phi+\Phi^\dagger {A}_\mu \Phi \partial^\mu \Phi^\dagger \Phi\right\}  -i \partial^\mu \left(\Phi^\dagger {A}_\mu \Phi\right) 
\notag\\
= & i \Phi^\dagger {A}_\mu \partial^\mu \Phi-i \partial^\mu \Phi^\dagger {A}_\mu \Phi-2i \Phi^\dagger {A}_\mu \Phi \Phi^\dagger \partial^\mu \Phi \ ,
\label{eq:DM_phi}
\end{align}
where we have used $\pd_\mu \Phi^\dagger \Phi = -\Phi^\dagger\pd_\mu \Phi$.
In addition, we have
\begin{equation}
    (\A_\mu\times\n)^2 = |\A_\mu|^2|\n|^2 - (\A_\mu\cdot\n)^2 = (A_\mu^a)^2 - (\Phi^\dagger A_\mu \Phi)^2 \ .
    \label{eq:A_second}
\end{equation}
We finally find from Eqs.~\eqref{eq:Lag_Phi_expand}, \eqref{eq:Dirichlet_term}, \eqref{eq:DM_phi} and \eqref{eq:A_second} that 
the $U(1) \times SU(2)$ gauged linear sigma model given in Eq.~\eqref{eq:U2_gauge_theory}
reduces in the strong coupling limit to the $SU(2)$ gauged $O(3)$ nonlinear sigma model 
in Eq.~\eqref{eq:gauged_sigma_model}.

\end{appendix}

\bibliographystyle{jhep}
\bibliography{references}

\end{document}